\documentclass[10pt,aps,prb,twocolumn,floatfix,floats,showpacs,superscriptaddress,raggedbottom]{revtex4-1}
\usepackage{epsfig}
\usepackage{graphicx,latexsym}
\usepackage{dcolumn}
\usepackage{amssymb,amsmath,bm}

\usepackage[dvips]{color}

\usepackage{hyperref}
\hypersetup{
    pdfnewwindow=true,       % links in new window
    colorlinks=true,         % false: boxed links; true: colored links
    linkcolor=blue,          % color of internal links
    citecolor=blue,          % color of links to bibliography
    filecolor=magenta,       % color of file links
    urlcolor=black           % color of external links
}

\def\eq#1{Eq.\ (\ref{#1})}
\def\mb#1{\mbox{\boldmath$#1$}}
\def\fig#1{Fig.\ \ref{#1}}

\begin{document}
\title{Finger-gate manipulated quantum transport in a semiconductor narrow constriction \\
 with spin-orbit interactions and Zeeman effect}
 \author{Chi-Shung Tang}
 \email{cstang@nuu.edu.tw}
 \affiliation{Department of Mechanical Engineering, National United University,
 Miaoli 36003, Taiwan, Republic of China}
 \author{Shu-Yu Chang}
 \affiliation{Department of Electrophysics, National Chiao Tung University,
 Hsinchu 30010, Taiwan, Republic of China}
 \author{Shun-Jen Cheng}
 \email{sjcheng@mail.nctu.edu.tw}
 \affiliation{Department of Electrophysics, National Chiao Tung University,
 Hsinchu 30010, Taiwan, Republic of China}

\date{\today }

\begin{abstract}
\ The authors investigate quantum transport in a narrow constriction
fabricated by narrow band gap semiconductor materials with
spin-orbit (SO) couplings. We consider the Rashba-Dresselhaus (RD)
spin-orbit interactions (SOIs) and the Zeeman effect induced by an
in-plane magnetic field along the transport direction.  The
interplay of the RD-SOI and the Zeeman effect may induce a
SOI-Zeeman gap and influence the transport properties.  We
demonstrate that an attractive scattering potential may induce
electron-like quasi-bound-state feature and manifest the
RD-SOI-Zeeman induced Fano line-shape in conductance. Furthermore, a
repulsive scattering potential may induce hole-like
quasi-bound-state feature on the subband top of the lower spin
branch.
\end{abstract}

%72.25.Dc Spin polarized transport in semiconductors
%72.30.+q High-frequency effects; plasma effects in electronic transport

\maketitle

\section{Introduction}

Quantum transport involving interference nature of charged particle
can be realized by using the split-gates induced narrow constriction
connecting the source and drain Ohmic contacts. The conductance
through the narrow constriction is known to be quantized when the
Fermi level of the system is tuned energetically by applying a
voltage to a nearby gate.\cite{Wees1988,Wharam1988} The quantization
features can be explained within the framework of simple
noninteracting models,\cite{Landauer1970,Buttiker1990,Beenakker1991}
and the conductance depends only on the transmission coefficient.
The related quantum devices can be utilized in various applications
including the prototypes of quantum information
processing.\cite{Loss1998}

Spin-orbit interaction (SOI) is a relativistic effect, in which a
charged particle moving with direction perpendicular to an electric
field experiences an effective magnetic field that couples to the
spin degree of freedom of the moving particle. Various spin-orbit
(SO) effects present in semiconductor structures provide a promising
way to spin manipulation in two-dimensional (2D) electron
gases.\cite{Winkler2003,Meier2007} Band structure behaviors and
transport properties involving SOI in semiconductor quantum
structures have received much interest due to its important
application in the emerging field of spintronic
devices.\cite{Datta1990,Zutic2004,Bandyopadhyay2004} Manipulating
the spin degree of freedom offers the possibility of devices with
high speed and very low power dissipation that is one of the
essential requirement for the applications in quantum computing and
memory storage.\cite{Wolf2001,Awschalom2002}

The SOI can be induced when the transporting electron experiences a
strong electric field due to the asymmetry in the confinement
potential, namely the structure inversion asymmetry (SIA) induced
Rashba SOI.\cite{Rashba60}  Especially, the Rashba SOI may be
significantly induced in two-dimensional electron gases (2DEGs)
confined by asymmetric potential in semiconductor materials.
Experimentally, the Rashba interaction has been shown to achieve
electron spin manipulation by using bias-controlled gate
contacts.\cite{Nitta1997}

In addition to the Rashba effect, there is also a Dresselhaus SOI
caused by the microscopic electric field arising from the lack of
inversion symmetry in the Bravais lattice, namely the bulk inversion
asymmetry (BIA).\cite{Dresselhaus1955}  The combined effect of the
Rashba and Dresselhaus SOI affects significantly the spin related
properties and should be considered when analyzing the performance
of spin-resolved electronic devices. Recently, several approaches
were proposed to engineer the spin-resolved subband structure
utilizing magnetic fields\cite{Muccio02,Brataas02,Zhang03,Wang03} or
ferromagnetic materials.\cite{Sun03,Zeng03}  The SOI and in-plane
magnetic field induced Zeeman effect may modify the subband
structure leading to SOI-Zeeman subband gap
feature.\cite{Pershin2004,Quay2010}  However, how the scattering
potentials influence the spin-resolved quantum transport and its
interplay with the SOI-Zeeman interactions has not yet been
explored.

In this work, we consider a  split-gate induced narrow constriction
that is fabricated in a 2D quantum well with narrow band gap
semiconductor material. Both the Rashba and Dresselhaus SOIs as well
as an applied external in-plane magnetic field are taken into
account to investigate the influences of the subband structures.
Moreover, we apply a narrow finger gate to affect the ballistic
transport properties. Below, we shall demonstrate analytically and
numerically that tuning the strength of the applied in-plane
magnetic field as well as the Rashba and the Dresselhaus SO-coupling
constants to manipulate the subband structures leading to fruitful
quantum transport properties.

This article is organized as follows. In Sec.\ II, we shall describe
our theoretical model including the Rashba and Dresselhaus SOIs as
well as an external in-plane magnetic field.  Section III
investigates the spin-resolved quantum transport properties.
Concluding remarks will be presented in Sec.\ IV.

\section{model and subband structures}

The system under investigation is assumed to be a narrow band gap
InAs-In$_{1-x}$Ga$_x$As semiconductor heterostructure grown in
$[0,0,1]$ crystallographic direction.  We consider the conduction
band of a 2D quantum well within the effective mass approximation.
We select the length unit $l^* = 1/k_F$ is the inverse of the Fermi
wave number $k_F$, and the energy unit  $E^* = E_F$ is the Fermi
energy $E_F = \hbar^2 k_F^2/2m^*$ with $m^\ast$ and $\hbar$ being,
respectively, the effective mass of an electron and the reduced
Planck constant. Correspondingly, the magnetic field is in units of
$B^* = E^*/\mu_B$ with $\mu_B$ being the Bohr magneton, and the
Rashba and Dresselhaus SO-coupling constants are in units of
${\alpha}^* = {\beta}^* = E^*l^*$. By using the above units, all
physical quantities presented below are
dimensionless.\cite{Tang1996}

\begin{figure}[tbhq]
      \includegraphics[width=0.45\textwidth,angle=0]{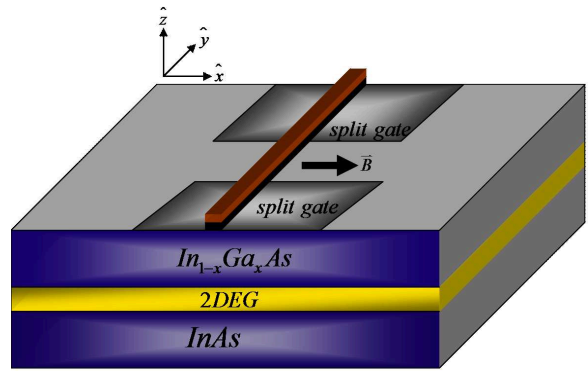}
      \caption{(Color online) Schematic illustration of the quantum channel
      defined by a pair of split gates that is fabricated by a narrow band gap
      InAs-In$_{1-x}$Ga$_x$As semiconductor heterostructure forming
      the two-dimensional electron gas (2DEG). An external in-plane
      magnetic field $\mathbf{B} = B \hat{\mb{x}}$ and a top finger
      gate are applied to influence the spin-resolved quantum transport
      properties.
      }
      \label{fig1}
\end{figure}

A pair of split gates is applied in the transverse direction forming
a quantum channel described by the unperturbed Hamiltonian
\begin{equation}
H_0 = k^2 + V_{\rm c}(y)
 \label{H0}
\end{equation}
that consists of a 2D kinetic energy term $k^2=k_x^2 + k_y^2$  and a
confining potential energy term
\begin{equation}
{V_c}(y) = \left\{ \begin{array}{l}
0, \, \left| y \right| < W/2\\
\infty, \, {\rm{otherwise}}
\end{array} \right.
\label{Vc}
\end{equation}
where $W$ indicates the width of the quantum channel.  The
transported electron is supposed to be affected by the effects of SO
interaction and the external in-plane magnetic field, and hence can
be described by the effective unperturbed Hamiltonian
\begin{equation}
\widetilde{H}_0 = H_0 + H_{\rm SO} + H_{\rm Z}. \label{H0til}
\end{equation}
Here we have assumed that the magnetic field is applied in the
transport direction $\hat{\mb{x}}$ such that the Zeeman interaction
is simply $H_{\rm Z}= gB \sigma_x$, in which the factor $g = g_s/2$
with $g_s$ being the effective gyromagnetic factor ($g_s = -15$ for
InAs). In order to manipulate the spin-resolved quantum transport
properties, we apply a finger gate on top of split gate with an
insulator in between, as illustrated in \fig{fig1}. We assume that
the finger gate is sufficiently narrow and then can be described by
a delta scattering potential form $V_{\rm sc}(x) = V_0 \delta(x)$.
The whole quantum channel system under investigation is thus
described by the total Hamiltonian $H = \widetilde{H}_0 + V_{\rm
sc}(x)$.

The SOI term in \eq{H0til} consists of the Rashba and Dresselhaus
SOI effects $H_{\rm SO}= H_{\rm R} + H_{\rm D}$. For the transport
direction $\hat{\mb{x}}
\parallel [1,0,0]$, the Rashba SO Hamiltonian is given by the
$k$-linear form
\begin{equation}
H_{\rm R} =  \alpha \left( \sigma_x k_y - \sigma_y k_x \right) ,
\end{equation}
where $\sigma_i$ ($i$ = \{$x,y,z$\}) are the Pauli matrices and
$\mathbf{k}=(k_x,k_y)$ is the 2D electron wave vector. The Rashba
coupling strength $\alpha$ is proportional to the electric field
along $\hat{\mb{z}}$ direction perpendicular to the 2D electron gas.
In general, the Dresselhaus interaction has a cubic dependence on
the momentum of the carriers. For a narrow semiconductor quantum
well grown along the $[0,0,1]$ direction, it reduces to a 2D linear
momentum dependent form
\begin{equation}
H_{\rm D} = \beta \left( \sigma_x k_x - \sigma_y k_y \right),
\end{equation}
where the Dresselhaus coupling strength $\beta$ is determined by the
semiconductor material and the geometry of the sample.  The
spin-orbit coupling contributions can be simplified as $H_{\rm SO} =
\left( - \alpha \sigma_y + \beta \sigma_x \right) k_x$ in a narrow
quantum channel.

The eigenfunction of \eq{H0til} can be expressed as the
multiplication of the spatial wave functions and the spinor state
$\chi_n$,
\begin{equation}
\Psi (x,y) = \sum_n {\phi _n}(y){e^{i{k_x}x}}\chi_n , \label{Psi}
\end{equation}
where the transverse wave function in the subband $n$ is of the form
\begin{equation}
{\phi _n}\left( y \right) = \sqrt {\frac{\pi }{W}} \sin
 \left( {\frac{n\pi}{W}y} \right)
 \label{phi}
\end{equation}
with unperturbed subband energy $\varepsilon_n = \left( n\pi /W
\right)^2$ due to the bare confining potential.  The corresponding
eigenenergies can be obtained

\subsection{Rashba-Zeeman effects}

In the absence of the Dresselhaus SOI, the Dresselhaus coupling
strength $\beta$ is identically zero.  In this subsection, we focus
on the the Rashba-Zeeman (RZ) effect, in which the spin resolved
subband energies can be obtained analytically~\cite{Wang2006}
\begin{equation}
E_n^{\sigma}  = {\varepsilon _n} + k_x^2 + \sigma \sqrt {{{(gB)}^2}
+ {{(2\alpha {k_x})}^2}} {\rm{  }} \label{En_RZ}
\end{equation}
and the spinor states
\begin{equation}
{\chi_n^\sigma} = \frac{1}{{\sqrt 2 }}\left[
{\begin{array}{*{20}{c}}
1\\
{\sigma {e^{i\theta (k_x)}}}
\end{array}} \right],
 \label{chiRZ}
\end{equation}
where $\sigma = \pm$ indicates the upper ($+$) and lower ($-$) spin
branches and $\theta (k_x) = {\tan^{-1}}\left( {2\alpha
{k_x}}/{\left| {gB} \right|} \right)$ describes the momentum
dependent spin orientation of an electron.  Defining the group
velocity of an electron in the $\sigma$ spin branch
\begin{equation}
v_g^{\sigma} = \frac{dE_n^{\sigma}}{d{k_x}} = 2{k_x} + \sigma
\frac{{4{\alpha^2}{k_{x}}}}{{\sqrt {{g^2}{B^2} + 4{\alpha ^2}k_x^2}
}}
 \label{vg}
\end{equation}
allows us to determine the local minimum (subband bottom) and local
maximum (subband top) in the subband structures by setting the group
velocity to be identically zero.

The calculations presented below are carried out under the
assumption that the electron effective mass $m^{\ast}=0.023 m_0$,
which is appropriate to the InAs-In$_{1-x}$Ga$_x$As semiconductor
interface with the typical electron density $n_e \sim
10^{12}$~cm$^{-2}$.\cite{Nitta1997}  Accordingly, the length unit is
$l^*$ = $1/k_F$ = 5.0~nm, the energy unit is $E^*$ = $E_F$ = 66~meV,
and the spin-orbit coupling parameters are in units of ${\alpha^*} =
{\beta^*} = 3.3 \times 10^{-10}$~eV$\cdot$m.  Below, we select the
width of the narrow constriction $W=\pi l^*=15.7$~nm so that the
unperturbed subband energy is simply $\varepsilon_n = n^2$.
Moreover, the range of the variation of energy $E$ is smaller than
the second unperturbed subband energy, namely $E < \varepsilon_2 E_F
= 4E_F$.  Furthermore, sufficient low temperature is required to
avoid thermal broadening effect, that is, $k_B T < 0.1 \Delta
\varepsilon$ (or $T < 23$~K).  We note in passing that the width of
the scattering potential $V_{\rm sc}(x)$ should be less than the
Fermi wave length $\lambda_F = 31.4$~nm to be described as a delta
potential.  We assume high-mobility semiconductor materials and,
hence, the impurities and defects can be neglected.

\begin{figure}[h]
\includegraphics[width=0.35\textwidth] {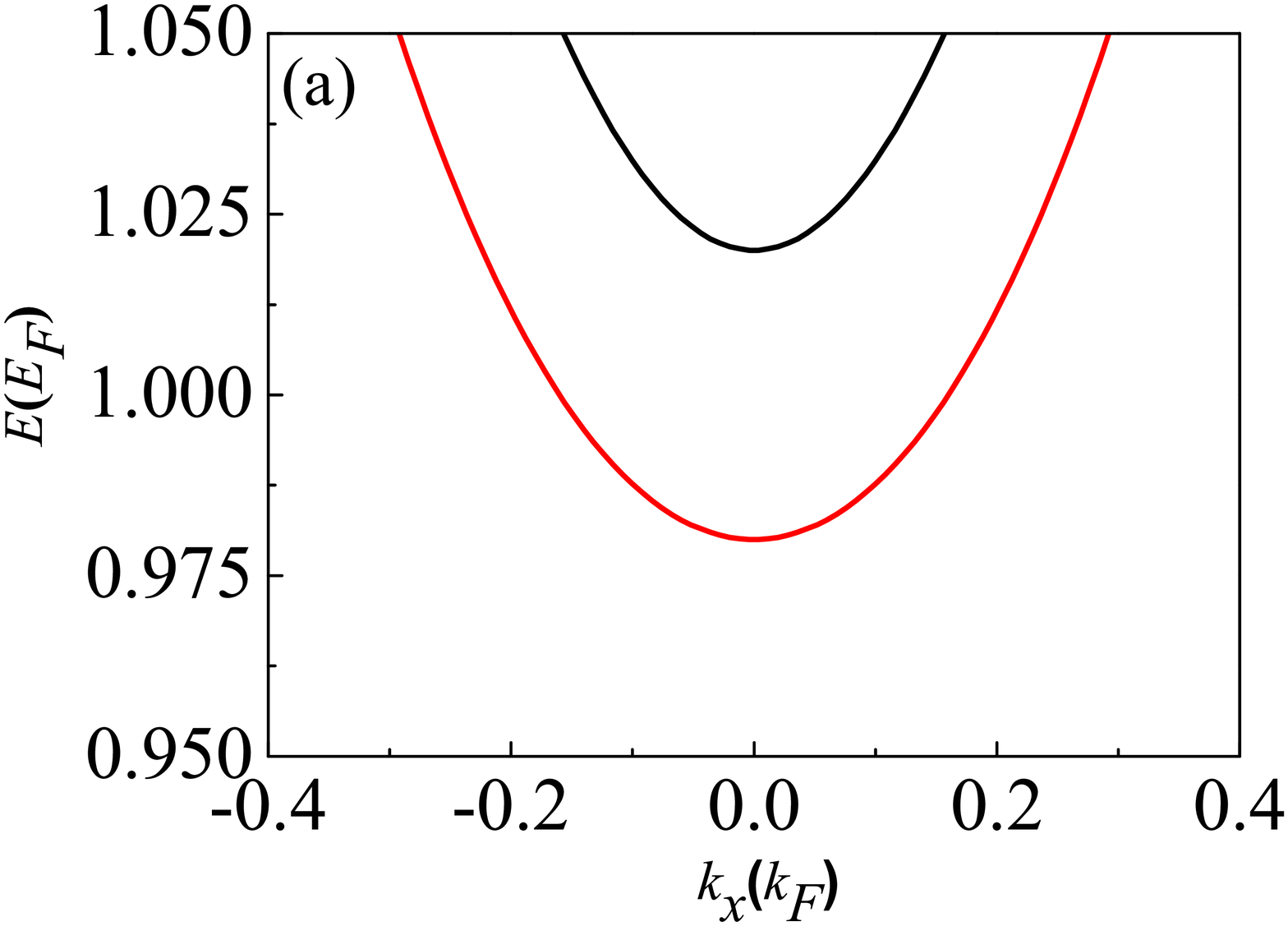}
\includegraphics[width=0.35\textwidth] {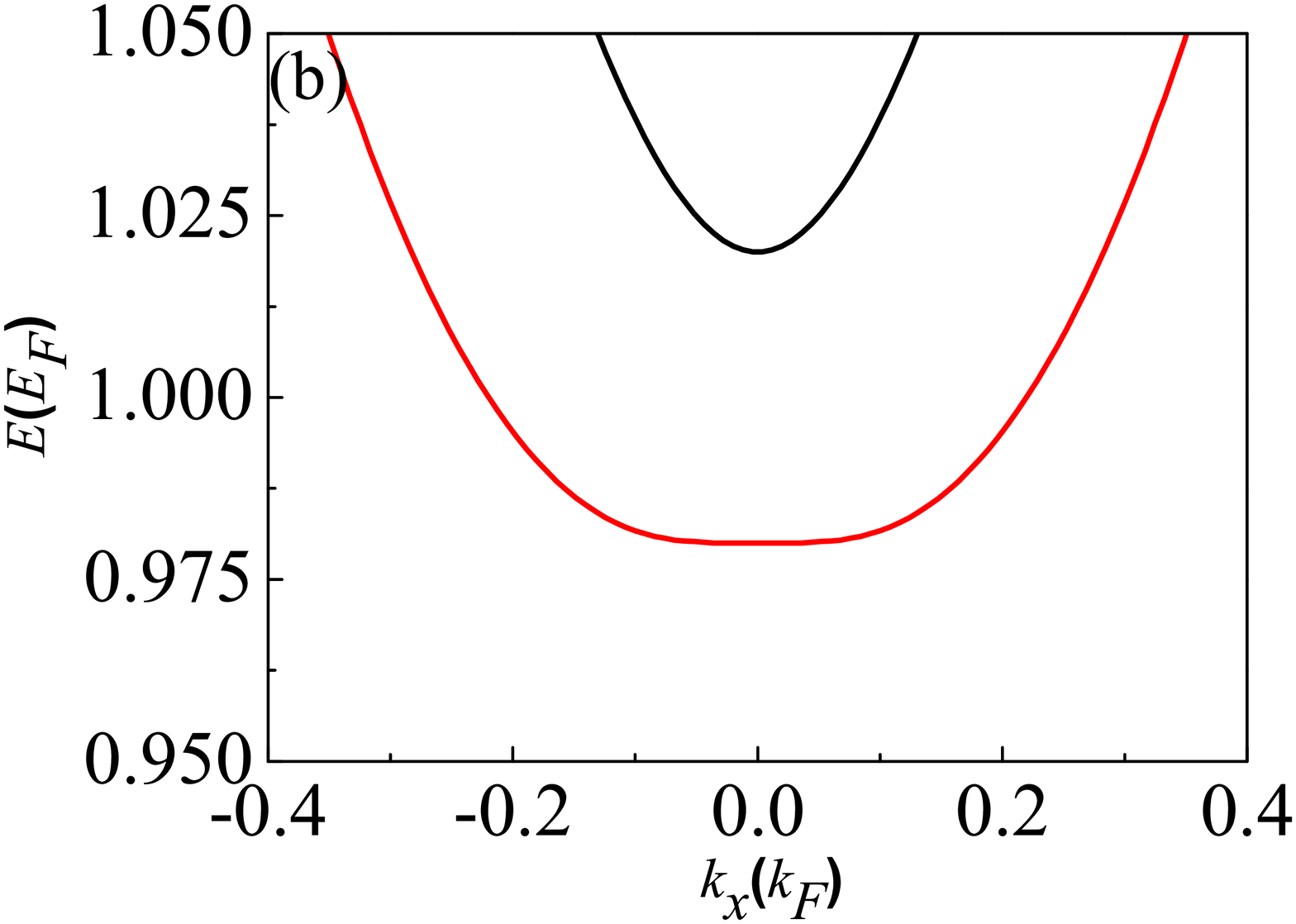}
\includegraphics[width=0.35\textwidth] {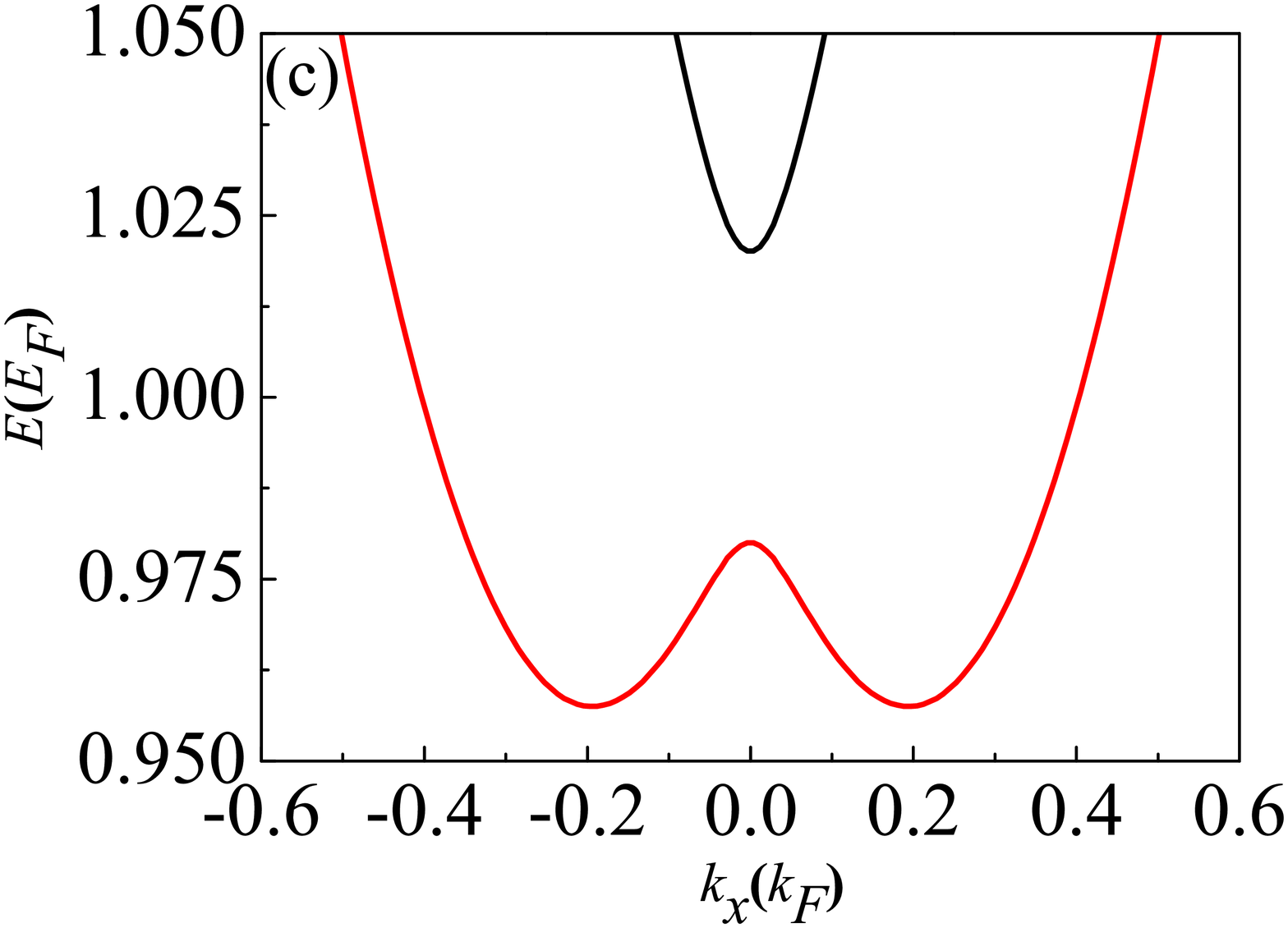}
\caption{(Color online) Energy spectrum versus wave number with
magnetic field strength $gB$ = 0.02 with different values of Rashba
coupling constant: (a) $\alpha$ = 0.05 (weak Rashba effect,
$2\alpha^2 < gB$); (b) $\alpha$ = 0.1 (critical Rashba effect,
$2\alpha^2 = gB$); and (c) $\alpha$ = 0.2 (strong Rashba effect,
$2\alpha^2 > gB$). The Fermi energy $E_F$ = 66 meV and the Fermi
wave number $k_F = 2\times 10^6 {\rm cm}^{-1}$. The magnetic field
strength is approximately 3T when $gB$ = 0.02 ($g_s = -15$ for
InAs). The black and red curves indicate the plus ($\sigma = +$) and
minus ($\sigma = -$) spin branches, respectively. } \label{E_RZ}
\end{figure}

The energy spectrum for the case of RZ effect with different
coupling strength regimes is illustrated in \fig{E_RZ}.  Before we
illustrate the subband gap features, it should be reminded that the
Zeeman effect is to induce an energy gap $\Delta E_{\rm{Z}} = 2gB$
between the opposite spin branches.   For the case of weak Rashba SO
coupling, namely $2\alpha^2 < gB$, both the spin branches have only
subband bottoms at $k_x = 0$ with energies at $E_n^\sigma =
\varepsilon_n +\sigma gB$.  The subband energy spacing between the
upper ($+$) branch and the lower ($-$) branch is the Zeeman
splitting $\Delta {E_{\rm{RZ}}}$ = $\Delta E_{\rm{Z}}$. Hence the
RZ-SO gap is dominated by the Zeeman effect in the weak SO coupling
regime.

For the case of strong Rashba coupling $2\alpha^2 > gB$, the subband
bottom of the upper spin branch is still at $E_n^+ = \varepsilon_n +
gB$. However, the subband bottom at $k_x = 0$ of the lower spin
branch becomes a subband top with the same energy $E_n^+ =
\varepsilon_n - gB$. Therefore, the subband energy spacing between
the $+$ and $-$ branches is still $\Delta {E_{\rm{Z}}} = 2gB$ but
forming a subband gap. In addition to the subband top in the lower
subband branch, there are two subband bottoms at $k_x = \pm \left[
\alpha^2 - (gB/2\alpha)^2 \right]^{1/2}$ with the same energy $E_n^-
= \varepsilon_n - \left[ \alpha^2 +(gB/2\alpha)^2\right]$.

We note in passing that if only the Rashba effect is considered, the
subband structure is simply $E_n^\sigma  = \varepsilon_n + k_x^2 +
\sigma 2\alpha {k_x}$.  Also, the subband structure manifests only
lateral splitting in momentum $\Delta k_x = 2\alpha$, where the
subband bottoms of $\sigma$ spin branches are at the wave numbers
$k_x^\sigma = -\sigma \alpha$ with the same energy $E_n =
\varepsilon_n - \alpha^2$.

In order to investigate the transport properties, one has to
determine the propagating and evanescent modes for a given energy.
To this end, it is convenient to rewrite the energy dispersion
relation in the following form
\begin{eqnarray}
\label{kx2RZ}
k_x^2 &=& (E + 2{\alpha ^2} - {\varepsilon _n}) \\
&\mp& \sqrt {{{(E + 2{\alpha ^2} - {\varepsilon _n})}^2} +
{{(gB)}^2} - {{(E - {\varepsilon _n})}^2}}. \nonumber
\end{eqnarray}
In general, this equation determines four complex $k_x$ values
corresponding to either propagating or evanescent modes.

\begin{figure}
\centerline{\includegraphics[width=8cm]{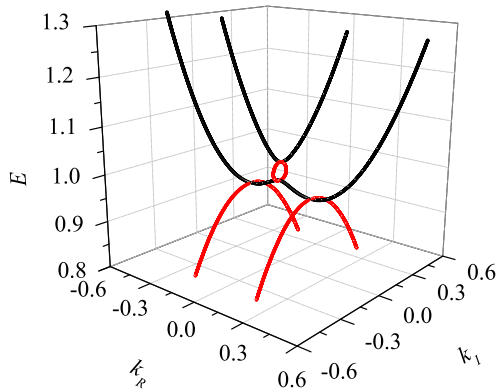}}
 \caption{(Color online) Energy (in
units of Fermi energy $E_F$) as a function of complex wave number
$k_x = k_R + i k_I$ (in units of Fermi wave number $k_F$). The
physical parameters are $\alpha$ = 0.2, $\beta$ = 0, $gB$ = 0.02.
The black curves represent the propagating modes and the red curves
denote the evanescent modes. }
 \label{E-krki}
\end{figure}
In \fig{E-krki}, we show the energy dispersion obtained from
\eq{kx2RZ} in the complex wave number space for the case of
$2{\alpha ^2} > gB$ so that subband gaps can be generated.  It is
clearly shown that there are four evanescent modes when the electron
energy is less than the lower subband bottom. When the electron
energy is greater than the lower subband bottom and below the
subband gap, there four propagating modes.  It is interesting to
notice that when electron energy is within the subband energy gap
regime, there are two propagating modes and two evanescent modes
(the red bubble in \fig{E-krki}). Although the conductance
calculated later only summing over the propagating modes, sufficient
number of evanescent modes should be taken into account to achieve
numerical accuracy when we calculate the intermediate scattering
processes.

\subsection{Rashba-Dresselhaus-Zeeman effects}

In the presence of Rashba and Dresselhaus SOI with an in-plane
magnetic field along the transport direction, the electronic system
can be described by
\begin{equation}
\varepsilon _n \Psi  + \left( {k_x ^2  - 2\alpha k_x \sigma _y  +
2\beta k_x \sigma _x  + gB\sigma _x } \right)\Psi  = E\Psi \, .
\label{RDZ}
\end{equation}
For electrons incident from the subband $n$, the spinor states
$\chi_n$ satisfies the $2\times 2$ matrix equation
\begin{eqnarray}
&& \left( {\begin{array}{*{20}c}
   {k_x^2 } & {gB + 2\beta k_x  + i2\alpha k_x }  \\
   {gB + 2\beta k_x  - i2\alpha k_x } & {k_x^2 }  \\
 \end{array} } \right)\chi_n \nonumber \\
 && = (E - \varepsilon _n )\chi_n \, .
\end{eqnarray}
The energy spectrum can be easily obtained of the form
\begin{equation}
 E = \varepsilon_n  + k_x^2  + \sigma
 \sqrt { (2 \beta k_x  + gB)^2 + (2 \alpha k_x )^2 }\, .
 \label{ERDZ}
\end{equation}
This equation is convenient to obtain energy spectrum as a function
of real wave vector for propagating modes.

\begin{figure}[h]
\includegraphics[width=6.0cm] {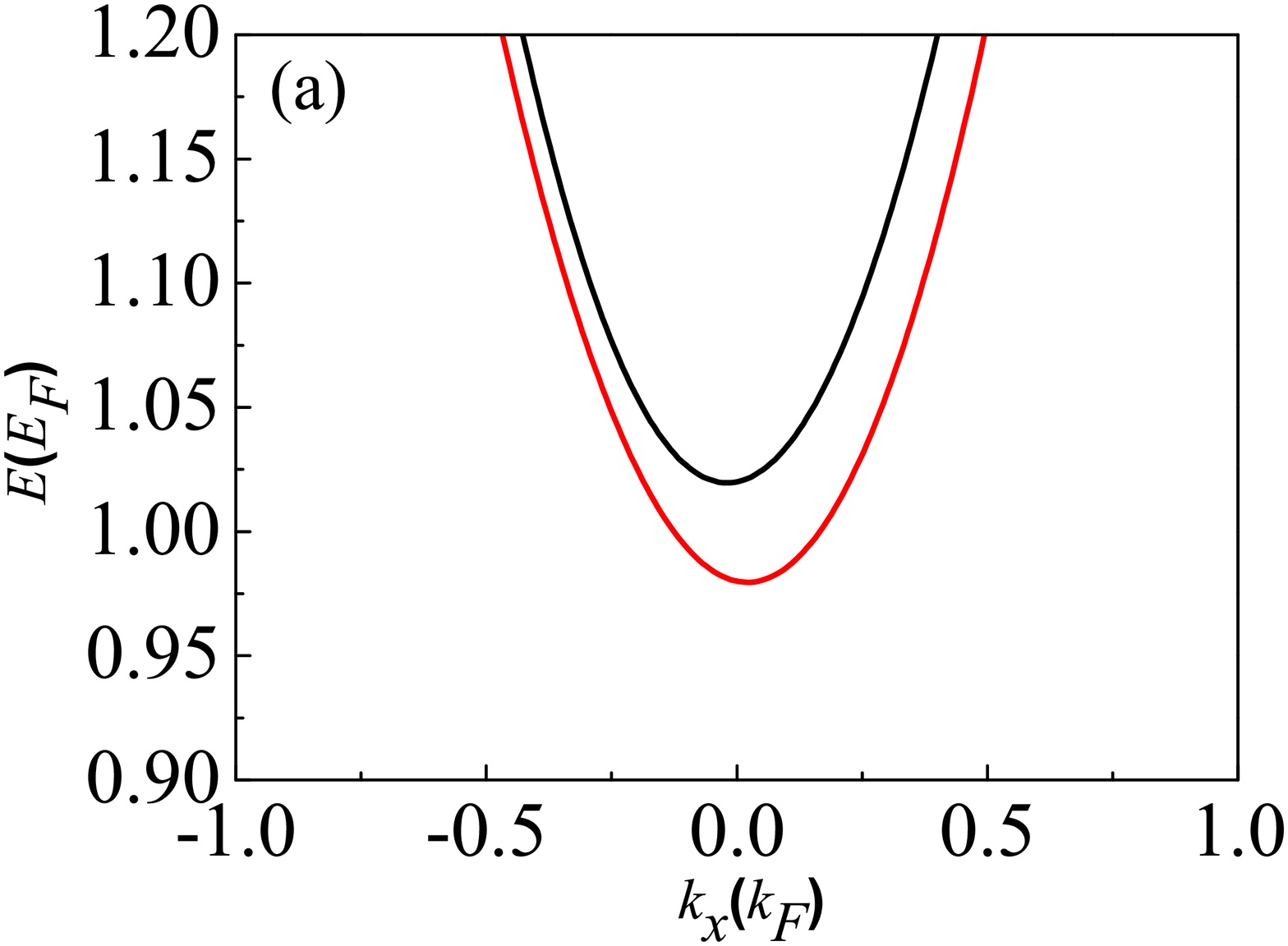}
\includegraphics[width=6.0cm] {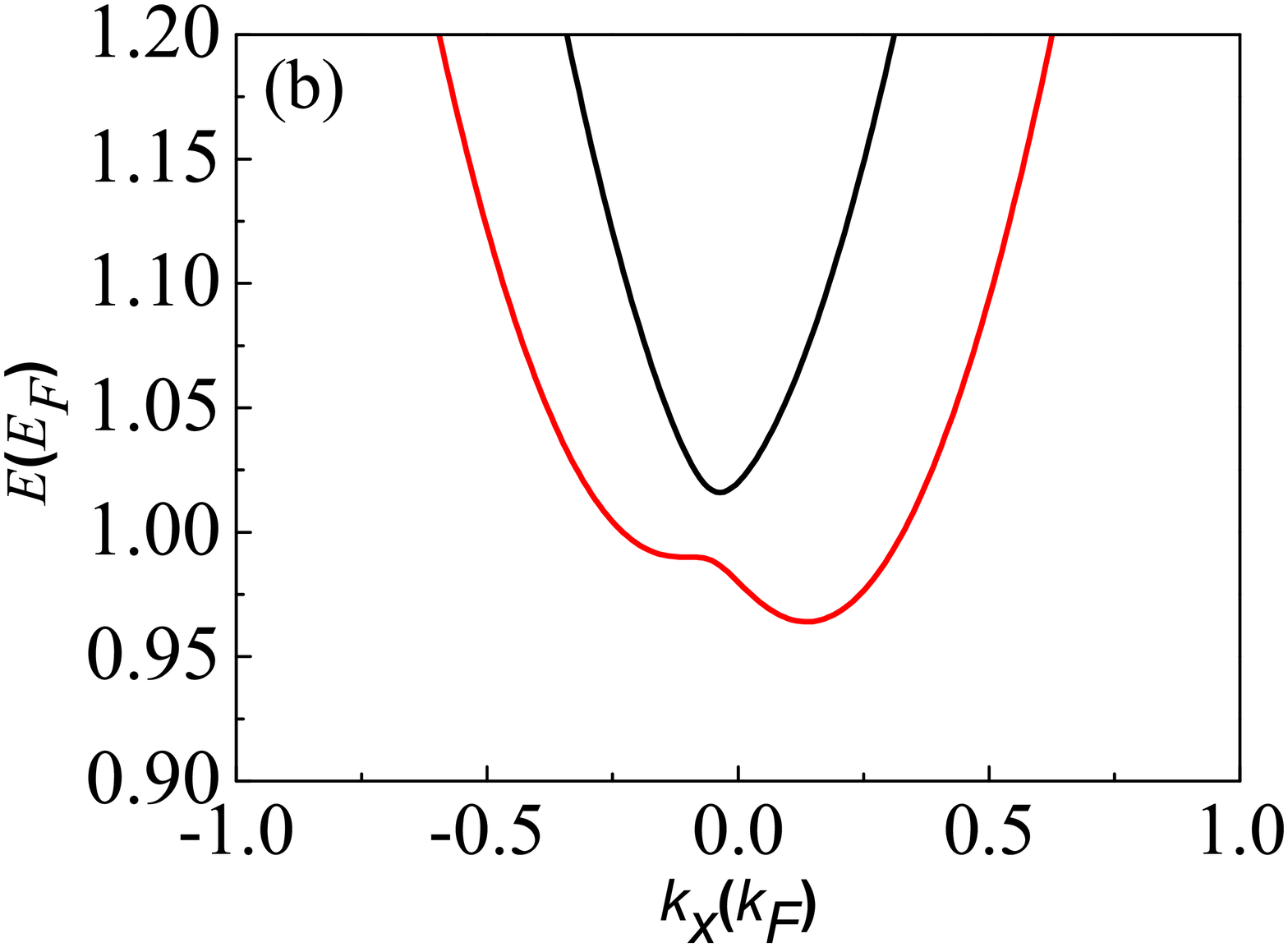}
\includegraphics[width=6.0cm] {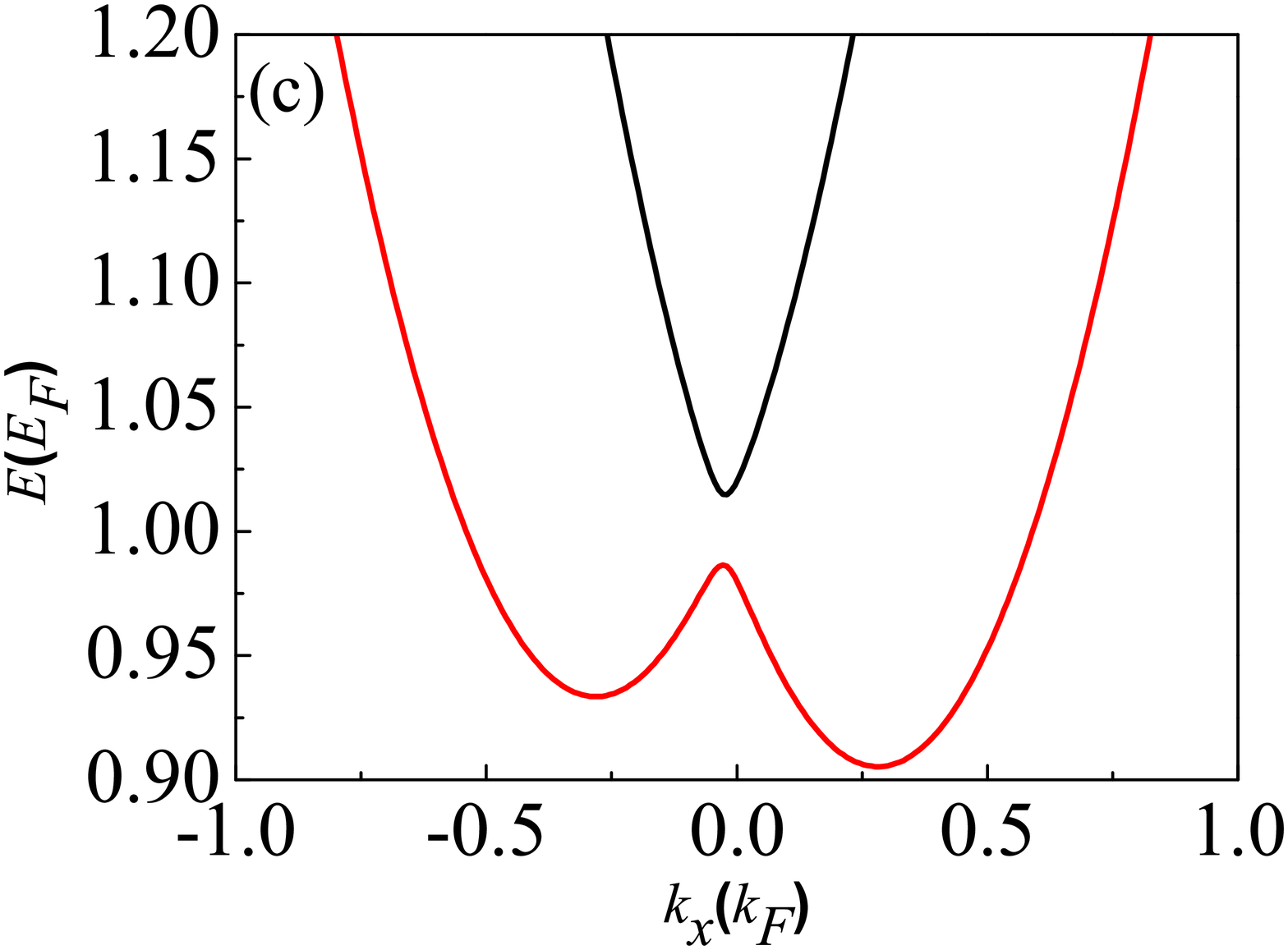}
\caption{(Color online) Energy spectrum versus wave number in the
presence of in-plane magnetic field ($gB$ = 0.02) with different
Rashba and Dresselhaus SO-coupling constants:  (a) $\alpha$ =
$\beta$ = 0.02 (weak coupling); (b) $\alpha$ = $\beta$ = 0.1
(mediate coupling); (c) $\alpha$ = $\beta$ = 0.2 (strong coupling).
The Fermi energy $E_F$ = 66 meV and the Fermi eave vector ${{k_F} =
{2 \times 10^6 {\rm cm}^{-1}} }$. The magnetic field strength is
approximately 3T when $gB $ = 0.02 ($g_s = -15$ for InAs). The black
and red curves indicate the plus ($\sigma = +$) and minus ($\sigma =
-$) spin branches, respectively.} \label{fig4.1.3.1}
\end{figure}

In \fig{fig4.1.3.1}, we show the Dresselhaus effect correction to
the Rashba SOI by fixing the strength of Zeeman effect ($gB$ = 0.02)
and tuning the strength of Rashba and Dresselhaus SOI.  To analyze
the the subband structures, it is convenient to introduce the
Rashba-Dresselhaus (RD) SO-coupling constant $\gamma$, namely
$\gamma^2 = \alpha^2 + \beta^2$, for defining different coupling
regimes. Figure \ref{fig4.1.3.1}(a) demonstrates the weak SO
coupling regime $\gamma^2 < gB$. It is clearly shown that the
spin-split subband structure is slightly asymmetric due to the
Dresselhaus effect. The subband bottoms of both subband branches are
no longer at zero wave number. Instead, the energy bottoms are
located at $(k_{x}, E_{1}^{+}) = (-0.02, 1.02)$ and $(k_{x},
E_{1}^{-}) = (0.02, 0.980)$. The two spin branches of a subband $n$
manifests a Zeeman splitting $\Delta E_{\rm Z}  = E_{n}^{+} -
E_{n}^{-} = 2gB = 0.04$. Hence, in the weak coupling regime, the
Zeeman effect dominates the subband structure and the RD coupling
slightly let the subband structure form an asymmetric lateral shift
in opposite direction for the spin branches.

Figure \ref{fig4.1.3.1}(b) illustrates the case of intermediate SO
coupling regime $\gamma^2 = gB$, it is shown that the lower spin
branch becomes a shoulder subband structure at $(k_{x}, E_{1}^{-}) =
(-0.1002, 0.99)$ and a clear subband bottom at $(k_{x}, E_{1}^{-}) =
(0.136, 0.964)$.  On the other hand, the subband bottom of the upper
branch is at $(k_{x}, E_{1}^{+}) = (-0.0361, 1.016)$, and hence the
spin branches form a shoulder gap feature $\Delta E_{sg} = 0.026 <
\Delta E_{\rm Z} $.

In Fig.\ \ref{fig4.1.3.1}(c), we show the case of strong SO coupling
regime ($\gamma^2 > gB$).  In this regime, it is interesting that
the lower spin branch manifests three extreme values in energy.
First, the left subband bottom of the upper spin branch is at
$(k_{x}, E_{1}^{-})$ = $(-0.28, 0.934)$. Second, the right subband
bottom is at $(k_{x}, E_{1}^{-})$ = $(0.28, 0.905)$. Third, the
subband top of the lower subband branch is at $(k_{x}, E_{1}^{-}) =
(-0.02, 0.986)$. On the other hand, the subband bottom of the upper
branch is at $(k_{x}^{-}, E_{1}^{-}) = (-0.02, 1.015)$. Therefore,
the subband gap in the strong coupling regime is around $\Delta E_g
= 0.029 > \Delta E_{sg}$. This implies that reduction of the subband
gap due to the Dresselhaus effect is a nontrivial effect.

\section{quantum transport properties}

In this section, we shall investigate the quantum transport
properties subject to spin-orbit interactions and Zeeman effect in a
narrow constriction. We assume that the quantum channel is
sufficiently narrow and focus on the first two conductance steps
associated with the two spin branches of an electron. Below, we
shall explore how the spin-mixing effect due to the SOI-Zeeman
coupling influences the transport properties.

\subsection{Rahsba-Zeeman effects}

\begin{figure}[bp]
\centerline{\includegraphics[width= 0.35 \textwidth]{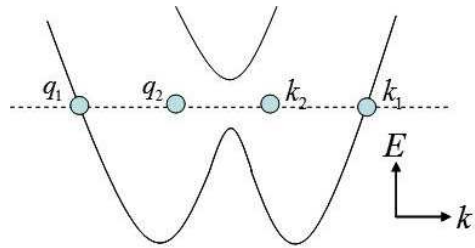}}
\caption{(Color online) Schematic illustration of the energy
spectrum with labeling modes. Here, $k_1$ and $q_1$ indicate the
outer right-going and outer left-going modes, respectively.
Similarly, $k_2$ and $q_2$ indicate the inner right-going and inner
left-going modes, respectively. In gap energy regime, both the inner
modes are evanescent modes. }\label{fig3.2.1}
\end{figure}

To consider an electron incident along the transport direction
$\mb{x}$, it is convenient to denote the wave number of right-going
(left-going) modes as $k_\sigma$ $(q_\sigma)$, in which the
subscript $\sigma$ could be by ``1" or ``2" indicating the ``outer"
or the ``inner" modes, as illustrated in \fig{fig3.2.1}. The
scattering wave function for an electron incident from the source
electrode can be written of the form
\begin{equation}
\psi (x) = {e^{i{k_\sigma }x}}\chi ({k_\sigma }) +
\sum\limits_\sigma ^{} {{r_\sigma }} {e^{i{q_\sigma }x}}\chi
({q_\sigma }),\, {\rm if} \, x < 0, \label{eq3.2.9}
\end{equation}
\begin{equation}
\psi (x) = \sum\limits_\sigma ^{} {{t_\sigma }{e^{i{k_\sigma
}x}}\chi ({k_\sigma }),\, {\rm if} \, x > 0.} \label{eq3.2.10}
\end{equation}
Here we have omitted the subband index for simplicity. By taking
into account the spin branches as well as the spin-flip scattering
mechanisms, the scattering wave functions can be generally expressed
as
\begin{eqnarray}
\psi (x) &=& {e^{i{k_\sigma }x}}\left[ {\begin{array}{*{20}{c}}
{{a_\sigma }}\\
{{b_\sigma }}
\end{array}} \right] + {r_\sigma }{e^{i{q_\sigma }x}}\left[ {\begin{array}{*{20}{c}}
{{c_\sigma }}\\
{{d_\sigma }}
\end{array}} \right] \nonumber \\
&&+ {r_{\bar \sigma }}{e^{i{q_{\bar \sigma }}x}}\left[
{\begin{array}{*{20}{c}}
{{c_{\bar \sigma }}}\\
{{d_{\bar \sigma }}}
\end{array}} \right],\, {\rm if} \, x > 0,
\label{eq3.2.11}
\end{eqnarray}
\begin{equation}
\psi (x) = {t_\sigma } \cdot {e^{i{k_\sigma }x}}\left\lfloor
{\begin{array}{*{20}{c}}
{{a_\sigma }}\\
{{b_\sigma }}
\end{array}} \right\rfloor  + {t_{\bar \sigma }}{e^{i{k_{\bar \sigma }}x}}
\left[ {\begin{array}{*{20}{c}}
{{a_{\bar \sigma }}}\\
{{b_{\bar \sigma }}}
\end{array}} \right], \, {\rm if} \, x> 0,
\label{eq3.2.12}
\end{equation}
where $ {a_\sigma } = 1{\sqrt 2}$ and ${b_\sigma } =  - {e^{i\theta
({k_\sigma })}} / {\sqrt 2 }$ with $\theta ({k_\sigma })$ = ${\tan
^{ - 1}}\left[ {{2\alpha {k_\sigma }}}/(gB) \right] $ are the spinor
elements of the incident wave. In addition, $ {c_\sigma } =
1/{\sqrt2}$ and ${d_\sigma}$ =  $-e^{ i\theta(q_\sigma)} / {\sqrt
2}$ with $\theta ({q_\sigma})$ = ${\tan^{-1}} \left[ 2\alpha
q_\sigma / (gB) \right]$ are the spin-state elements of the
reflection wave.  Moreover, the spin-state elements of the spin-flip
transmission wave are given by
\begin{equation}
{a_{\bar \sigma }} = \left\{ { \frac{ {{{(gB + 2\alpha {k_{\bar
\sigma }})}^2}}} {{{{(gB + 2\alpha {k_{\bar \sigma }})}^2} + \left|
{g^2B^2 - 4{\alpha ^2} k_{\bar \sigma }^2} \right|}}} \right\}^{1/2}
\label{eq3.2.19}
\end{equation}
and
\begin{equation}
{b_{\bar \sigma }} = {a_{\bar \sigma }} \cdot \frac{{\sqrt
{{g^2}{B^2} - 4{\alpha ^2} k_{\bar \sigma }^2} }}{{gB + 2\alpha
{k_{\bar \sigma }}}}\, . \label{eq3.2.20}
\end{equation}
Similarly, we can obtain the spin-state elements of the spin-flip
reflection wave, given by
\begin{equation}
{c_{\bar \sigma }} = \left\{ {\frac{{{{(gB + 2\alpha {q_{\bar \sigma
}})}^2}}}{{{{(gB + 2\alpha {q_{\bar \sigma }})}^2} + \left|
{{g^2}{B^2} - 4{\alpha ^2}q_{\bar \sigma }^2} \right|}}}
\right\}^{1/2} \label{eq3.2.21}
\end{equation}
and
\begin{equation}
{d_{\bar \sigma }} = {c_{\bar \sigma }} \cdot \frac{{\sqrt
{{g^2}{B^2} - 4{\alpha ^2} q_{\bar \sigma }^2} }}{{gB + 2\alpha
{q_{\bar \sigma }}}}\, . \label{eq3.2.22}
\end{equation}

By matching the boundary conditions at around the scattering
potential induced by the finger gate, it is easy to obtain the
matrix equation of motion for the spin-resolved transport involving
the finger-gate induced spin-flit scattering
\begin{eqnarray}
&& \left[ {\begin{array}{*{20}{c}}
{ - {c_\sigma }}&{ - {c_{\bar \sigma }}}&{{a_\sigma }}&{{a_{\bar \sigma }}}\\
{ - {d_\sigma }}&{ - {d_{\bar \sigma }}}&{{b_\sigma }}&{{b_{\bar \sigma }}}\\
{ - {q_\sigma }{c_\sigma }}&{ - {q_{\bar \sigma }}{c_{\bar \sigma }}}&{({k_\sigma }
+ i{V_0}){a_\sigma }}&{({k_{\bar \sigma }} + i{V_0}){a_{\bar \sigma }}}\\
{ - {q_\sigma }{d_\sigma }}&{ - {q_{\bar \sigma }}{d_{\bar \sigma
}}}&{({k_\sigma } + i{V_0}){b_\sigma }}&{({k_{\bar \sigma }} +
i{V_0}){b_{\bar \sigma }}}
\end{array}} \right]\left[ {\begin{array}{*{20}{c}}
{{r_\sigma }}\\
{{r_{\bar \sigma }}}\\
{{t_\sigma }}\\
{{t_{\bar \sigma }}}
\end{array}} \right] \nonumber \\
&&= \left[ {\begin{array}{*{20}{c}}
{{a_\sigma }}\\
{{b_\sigma }}\\
{{k_\sigma }{a_\sigma }}\\
{{k_\sigma }{b_\sigma }}
\end{array}} \right]\quad
\label{eq3.2.29}
\end{eqnarray}

To calculate conductance in the noninteracting electron model we
employ the framework of Landauer-B\"{u}ttiker
formula.\cite{Landauer1970,Buttiker1990}  For a given energy,
solving for the spin non-flip and flip reflection coefficients
$r_{\sigma }$ and $r_{\bar \sigma }$, as well as the spin non-flip
and flip transmission coefficients $t_{\sigma }$ and $t_{\bar \sigma
}$, we can thus express the zero temperature conductance as
\begin{equation}
G = G_0 \sum\limits_{\sigma_L,\sigma_R}
 \frac{v_{\sigma_R}}{v_{\sigma_L}}
 \left| t_{\sigma_L,\sigma_R} \right|^2 \, .
\label{eq3.2.30}
\end{equation}
Here $G_0$ = $e^2/h$ is the conductance quantum of a single spin
branch, and $\sigma_L$ and $\sigma_R$ indicate, respectively, the
spin branches of the incident and transmission waves in the left and
right leads. Therefore, ${v_{{\sigma _R}}}$ and ${v_{{\sigma _L}}}$
represent the group velocity at the corresponding modes.

\begin{figure}[h]
\includegraphics[width=0.4\textwidth,angle=0] {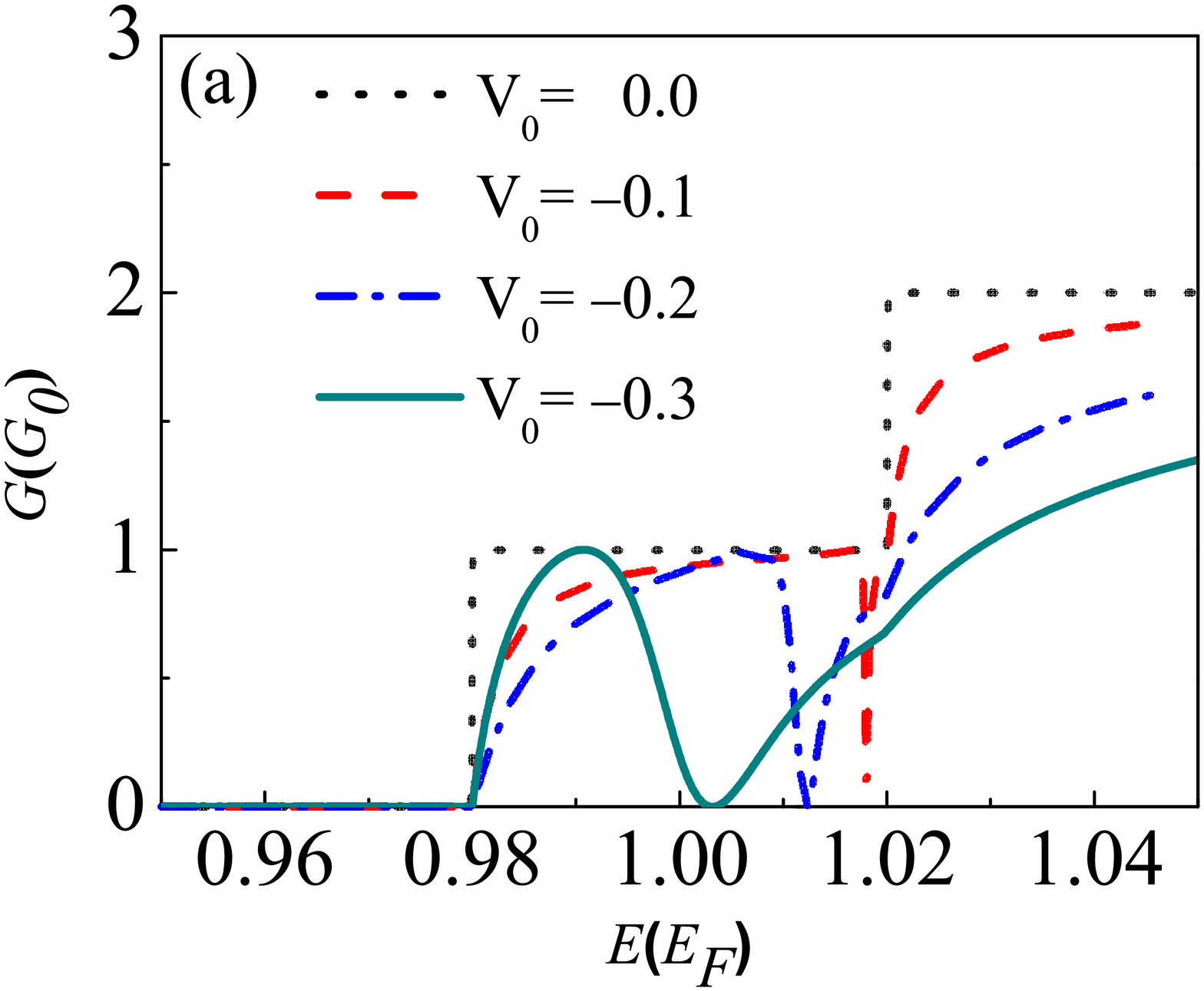}
\includegraphics[width=0.4\textwidth,angle=0] {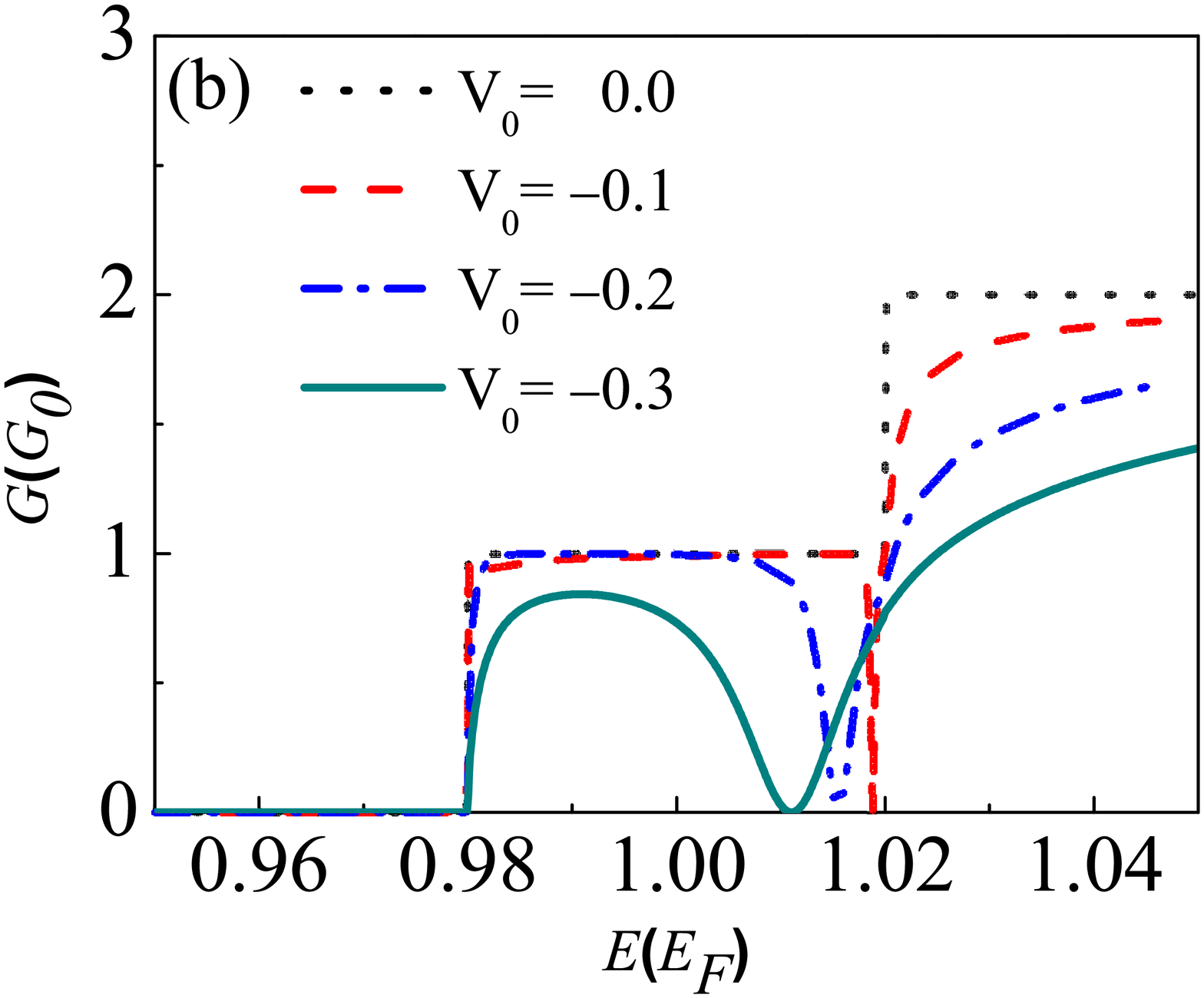}
\includegraphics[width=0.4\textwidth,angle=0] {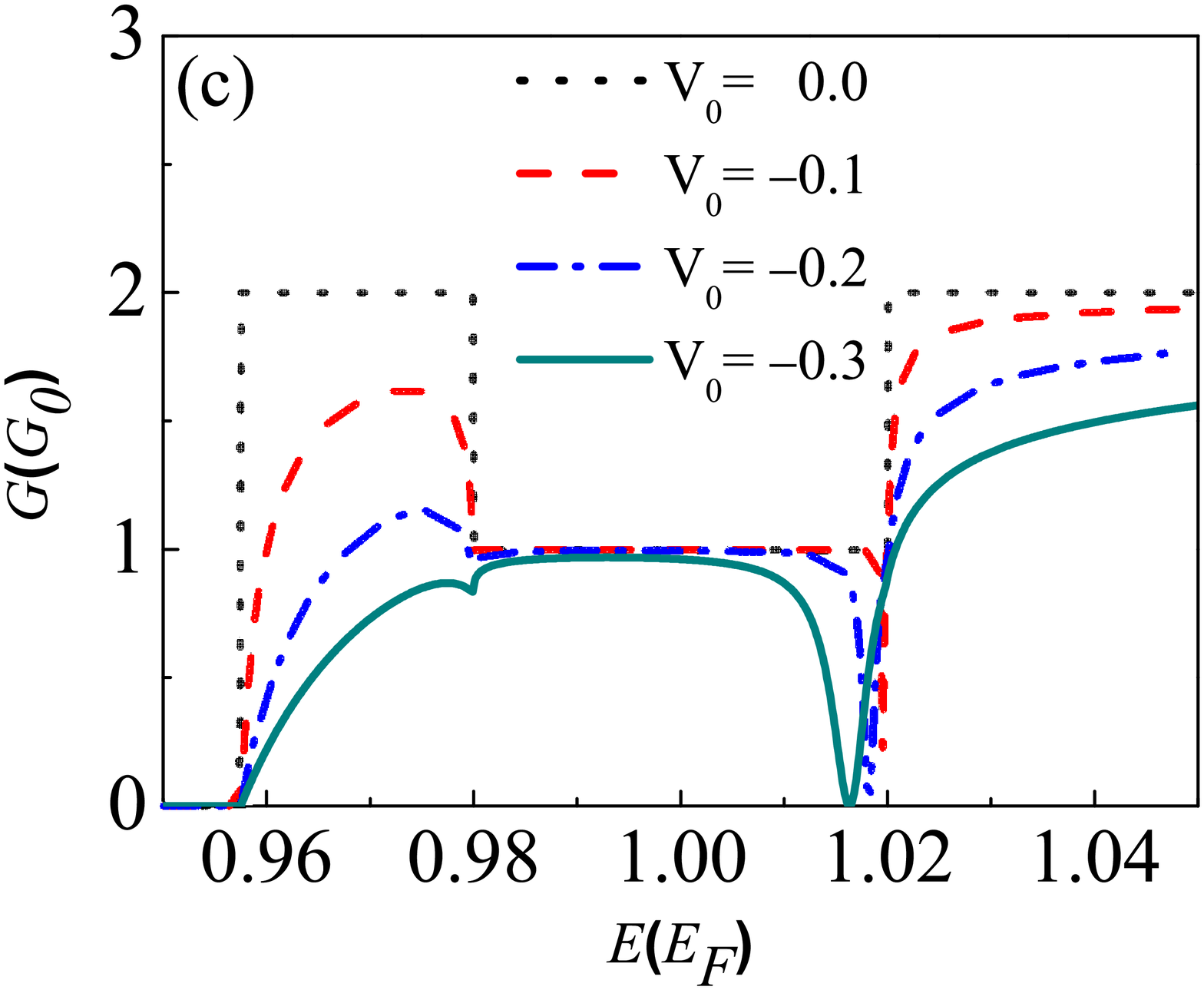}
\caption{(Color online) Conductance as a function of electron energy
with different strength of attractive potential: $V_0 = 0.0$ (dotted
black), $V_0 = -0.1$ (dashed red), $V_0 = -0.2$ (dash-dotted blue),
$V_0 = -0.3$ (solid green). We fix the Zeeman effect $gB$ = 0.02 (or
$B =3$~T if $g_s$ = $-15$ for InAs-based material) while tune the
Rashba parameter: (a) $\alpha$ = 0.05 ($2\alpha^2 < gB$, weak SO
coupling); (b) $\alpha$ = 0.1 ($2\alpha^2 = gB$, intermediate SO
coupling); (c) $\alpha$ = 0.2 ($2\alpha^2 > gB$, strong SO
coupling).} \label{fig3.3.2.1}
\end{figure}

In \fig{fig3.3.2.1}, we demonstrate the transport properties in the
presence of an attractive scattering potential due to the finger
gate by fixing the in-plane magnetic field ($gB$ = 0.02) while
tuning the strength of Rashba SOI. In general, the attractive
scattering potential plays a role to suppress the conductance.  we
present the conductance as a function of electron energy with
different strength of attractive potential: $V_0 = 0.0$ (dotted),
$V_0 = -0.1$ (dashed), $V_0 = -0.2$ (dash-dotted), $V_0 = -0.3$
(solid). Here, we fix the Zeeman effect to be $gB$ = 0.02, in other
words the magnetic field $B =3$~T if the factor $g_s$ = $-15$ for
InAs-based material. Moreover,  we tune the Rashba parameter: (a)
$\alpha$ = 0.05 ($2\alpha^2 < gB$, weak SO coupling); (b) $\alpha$ =
0.1 ($2\alpha^2 = gB$, intermediate SO coupling); (c) $\alpha$ = 0.2
($2\alpha^2 > gB$, strong SO coupling).

Figure \ref{fig3.3.2.1}(a) illustrates the transport properties in
the weak SO coupling regime ($2\alpha^2 < gB$). When the attractive
potential is weak ($V_0 = -0.1$), the conductance manifests a clear
dip structure and form an electron-like quasibound state at the
subband bottom of the upper spin branch. When the potential strength
increases ($V_0 = -0.3$), the dip structure becomes a broad valley
structure and is shifted toward the low energy regime indicating the
shorter life time. This broadening effect is suppressed in the
mediate SO coupling regime ($2\alpha^2 = gB$), as shown in
\fig{fig3.3.2.1}(b).  It is interesting to notice in
\fig{fig3.3.2.1}(c) that the conductance manifests an abrupt drop to
unity in the energy regime $0.98 < E < 1.02$ due to the spin-gap
feature as shown previously in \fig{E_RZ}(c).  It is interesting
that the conductance dip structure is not broadened for larger
scattering potential $V_0$. This indicates that the life time of the
quasibound state feature is enhanced in the strong Rashba SO
coupling regime.

\begin{figure}[h]
\includegraphics[width=0.4\textwidth,angle=0] {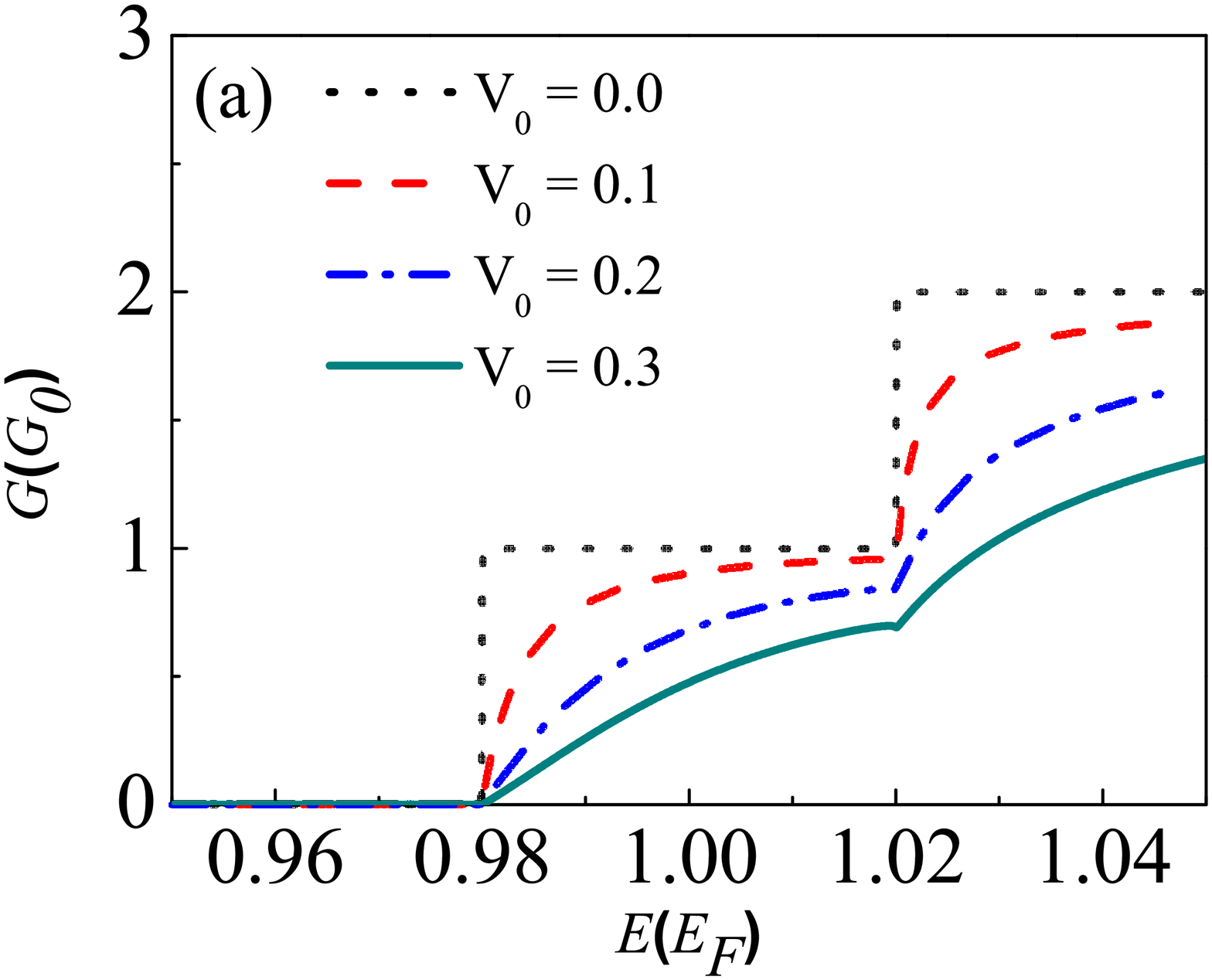}
\includegraphics[width=0.4\textwidth,angle=0] {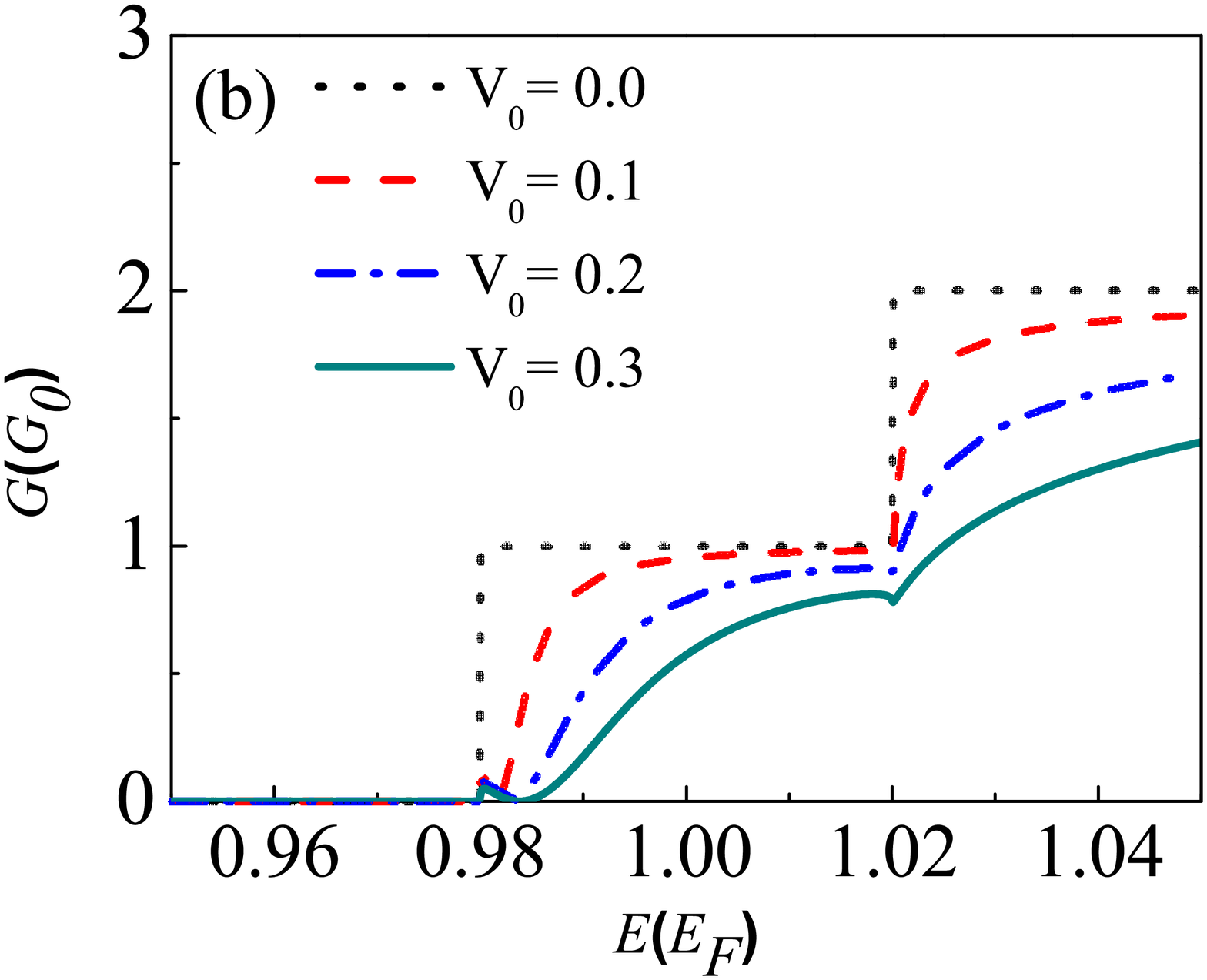}
\includegraphics[width=0.4\textwidth,angle=0] {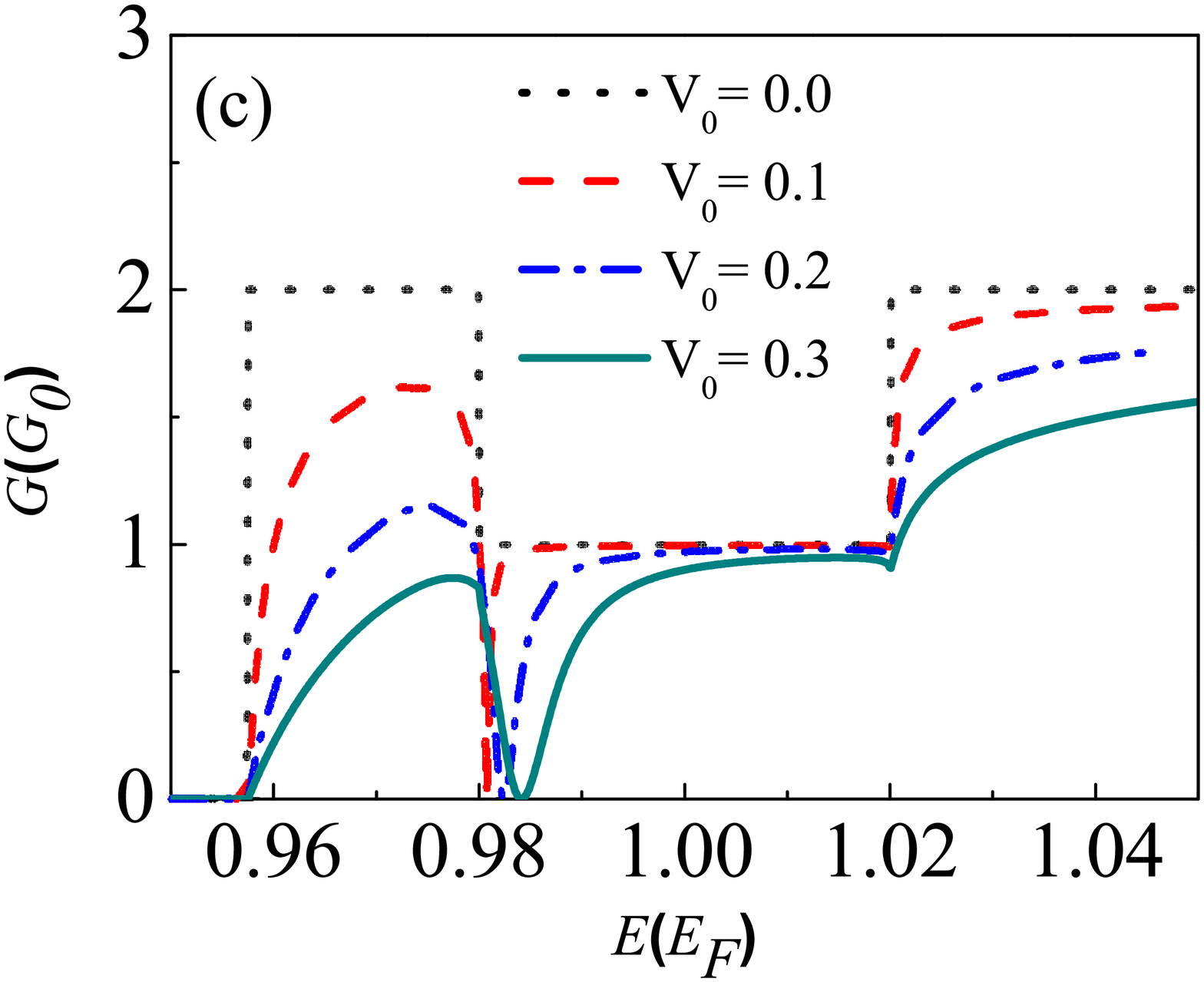}
\caption{(Color online) Conductance as a function of electron energy
with different strength of repulsive scattering potential: $V_0 =
0.0$ (dotted black), $V_0 = 0.1$ (dashed red), $V_0 = 0.2$
(dash-dotted blue), $V_0 = 0.3$ (solid green). We fix the Zeeman
effect $gB$ = 0.02 and tune the parameter of Rashba SOI: (a)
$\alpha$ = 0.05 ($2\alpha^2 < gB$, weak SO coupling); (b) $\alpha$ =
0.1 ($2\alpha^2 = gB$, intermediate SO coupling); (c) $\alpha$ = 0.2
($2\alpha^2 > gB$, strong SO coupling).} \label{fig3.3.2.2}
\end{figure}

In \fig{fig3.3.2.2}, we demonstrate the transport properties in the
presence of a repulsive scattering potential due to the finger gate
by fixing the in-plane magnetic field ($gB$ = 0.02) while tuning the
strength of Rashba SOI. In general, the repulsive potential plays a
role to strongly suppress the conductance in the low kinetic energy
regime. we exhibit the conductance as a function of electron energy
with different strength of repulsive scattering potential: $V_0 =
0.0$ (dotted), $V_0 = 0.1$ (dashed), $V_0 = 0.2$ (dash-dotted), $V_0
= 0.3$ (solid). Here, we fix the Zeeman effect to be $gB$ = 0.02,
namely the magnetic field $B =3$~T if the factor $g_s$ = $-15$ for
InAs-based material.  Then we tune the strength of the Rashba SOI:
(a) $\alpha$ = 0.05 ($2\alpha^2 < gB$, weak SO coupling); (b)
$\alpha$ = 0.1 ($2\alpha^2 = gB$, intermediate SO coupling); (c)
$\alpha$ = 0.2 ($2\alpha^2 > gB$, strong SO coupling).

For the case of weak SO coupling regime shown in
\fig{fig3.3.2.2}(a), the conductance is strongly suppressed in the
low kinetic energy regime and behaves monotonically increasing. For
the case of intermediate SO coupling regime shown in
\fig{fig3.3.2.2}(b), the conductance is more strongly suppressed in
the low kinetic energy regime than the case of weak SO coupling
regime. It is interesting to notice when the repulsive is strong
enough ($V_0 = 0.3$) the conductance is even suppressed to zero at
energy $E\approx 0.984 E^\ast$. This is a clue of hole-like
quasi-bound-state feature with very short life time due to the
shoulder-like structure of the lower subband branch shown in
\fig{fig3.3.2.2}(b).  For the case of strong SO coupling regime,
since the subband structure can form a subband gap, as is shown in
\fig{fig3.3.2.2}(c), it allows to form a significant hole-like
quasi-bound-state feature at the subband top of the lower spin
branch.  The conductance thus manifests a dip structure energy
$E\approx 0.984 E^\ast$.

\subsection{Rashba-Dresselhaus-Zeeman effects}

In this subsection we shall explore the transport properties of a
narrow constriction by fixing the Zeeman effect while manipulating
the strength of the RD-SOI and tuning the amplitude of scattering
potential that can be either attractive or repulsive.  All the
physical parameters shown bellow are the same with the physical
parameters in the previous subsection discussing the case of R-SOI.
The transport calculation for the case of RD-SOI is similar to the
case of R-SOI but has to be solved numerically not shown here.

\begin{figure}[h]
\includegraphics[width=0.4\textwidth,angle=0] {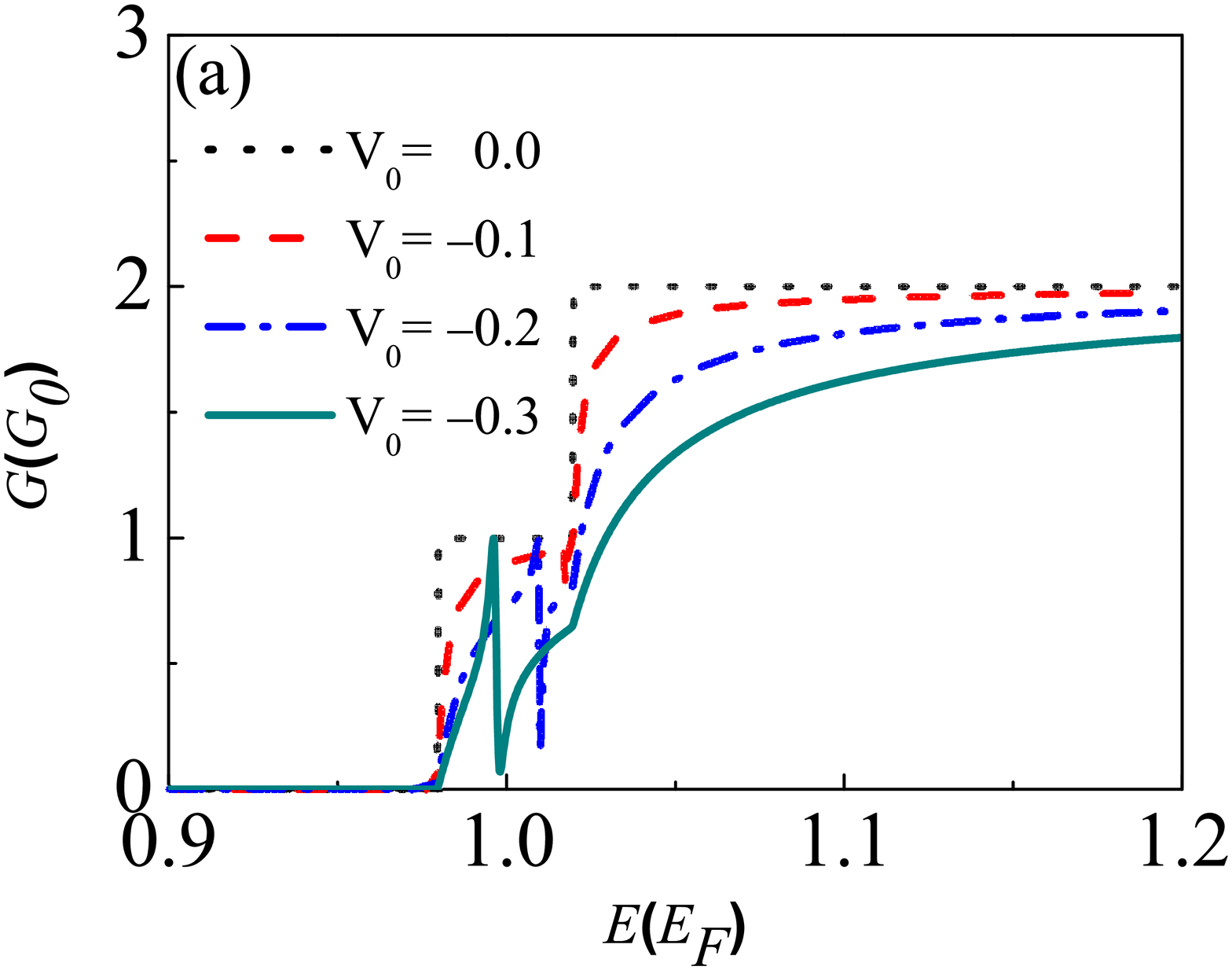}
\includegraphics[width=0.4\textwidth,angle=0] {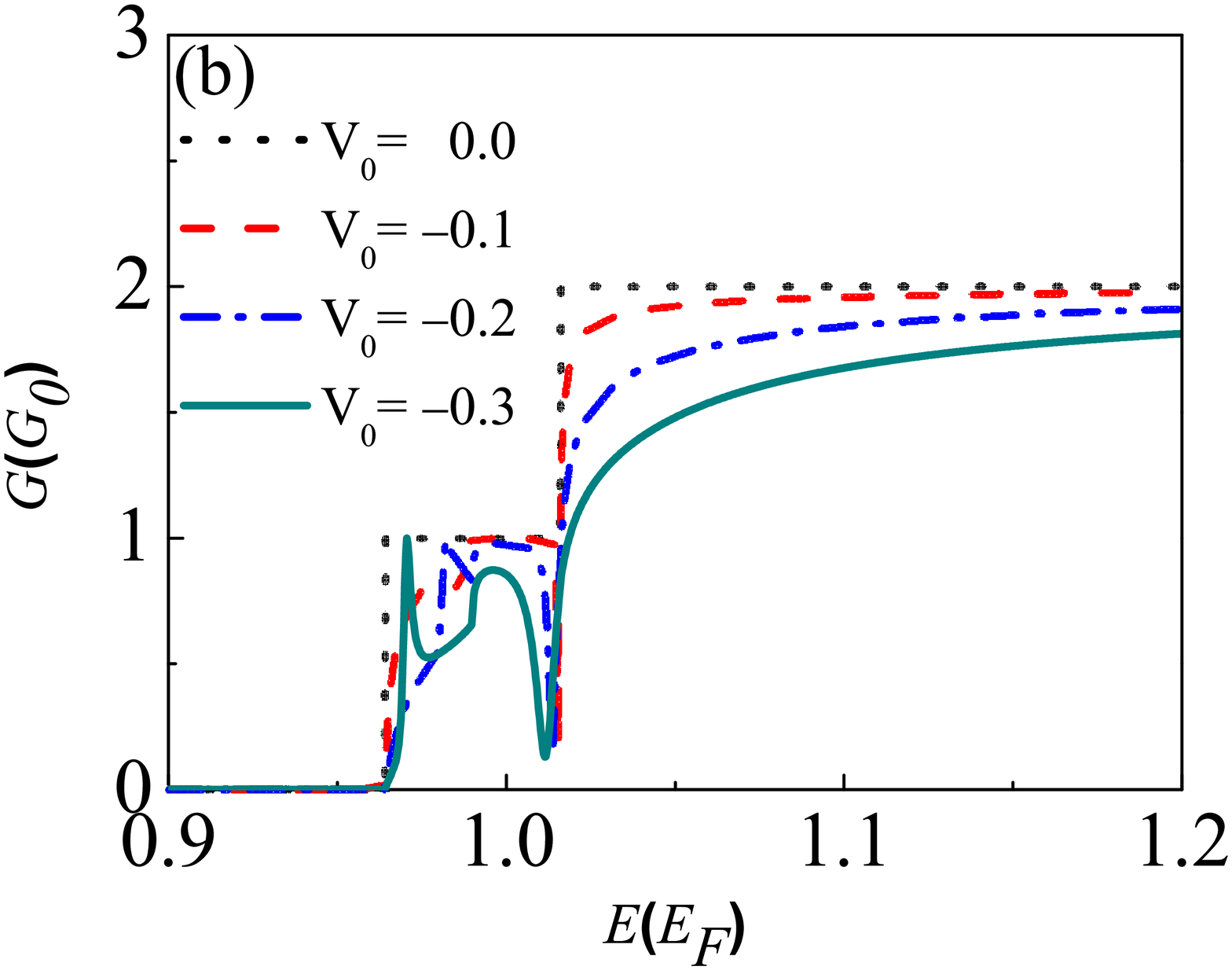}
\includegraphics[width=0.4\textwidth,angle=0] {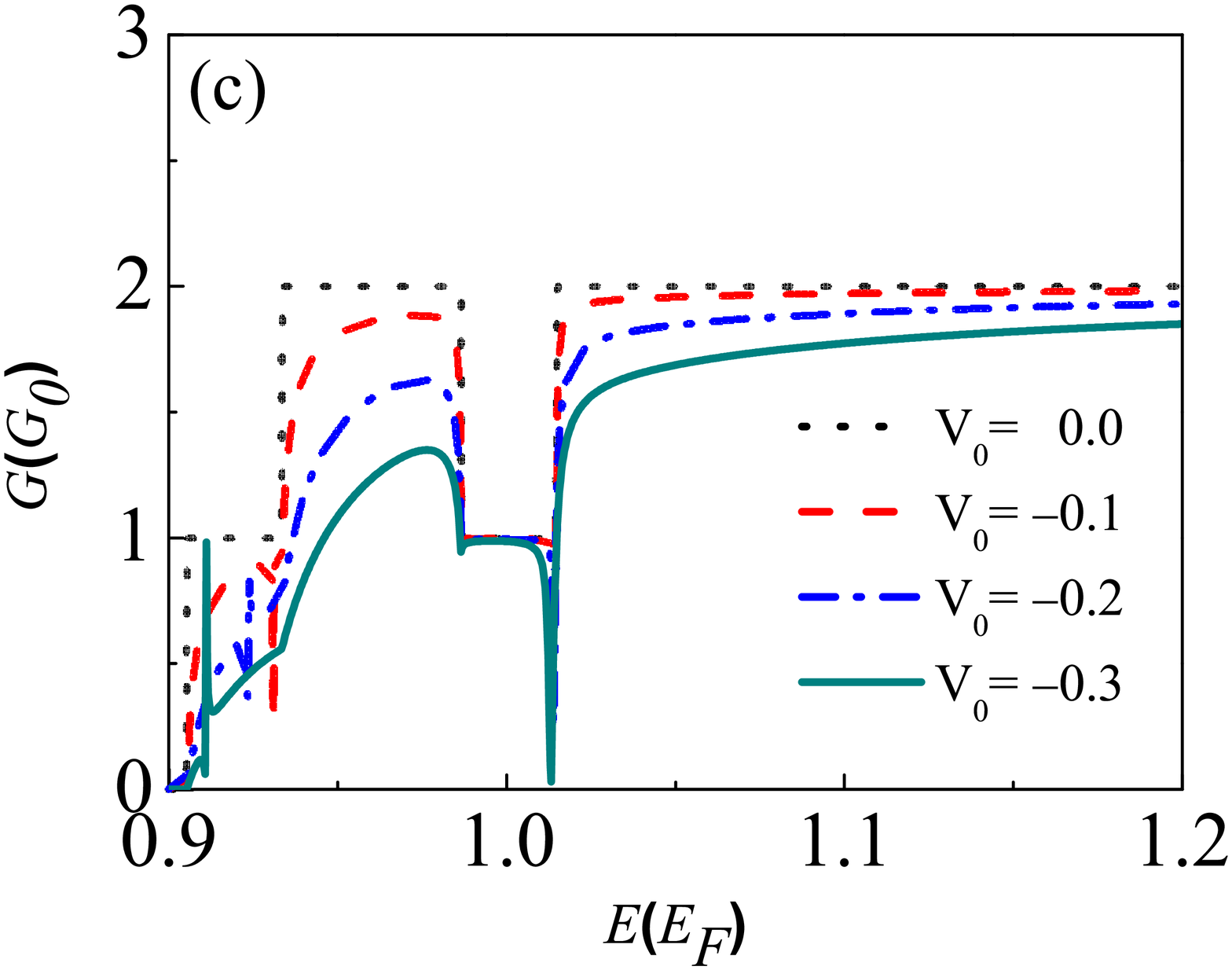}
\caption{(Color online) Conductance as a function of electron energy
with different strength of attractive scattering potential: $V_0 =
0.0$ (dotted black), $V_0 = -0.1$ (dashed red), $V_0 = -0.2$
(dash-dotted blue), $V_0 = -0.3$ (solid green). We fix the in-plane
magnetic field ($gB$ = 0.02) and tune the Rashba and Dresselhaus
SO-coupling constants: (a) $\alpha = \beta$ = 0.02 ($\gamma^2 < gB$,
weak SO coupling regime); (b) $\alpha = \beta$ = 0.1 ($\gamma^2 =
gB$, intermediate SO coupling regime); (c) $\alpha = \beta$ = 0.2
($\gamma^2 > gB$, strong SO coupling regime).} \label{fig4.3.4.1}
\end{figure}

In \fig{fig4.3.4.1}, we investigate how an attractive scattering
potential influences the transport properties by tuning Rashba and
the Dresselhaus effects and fixing the in-plane magnetic field, the
corresponding energy spectra are shown in \fig{fig4.1.3.1}. The
conductance is presented as a function of electron energy with
different strength of attractive scattering potential: $V_0 = 0.0$
(dotted), $V_0 = -0.1$ (dashed), $V_0 = -0.2$ (dash-dotted), $V_0 =
-0.3$ (solid). We fix the in-plane magnetic field so that the Zeeman
effect $gB$ = 0.02. In addition, the Rashba and Dresselhaus
SO-coupling constants are selected to cover three coupling regimes:
(a) $\alpha = \beta$ = 0.02 ($\gamma^2 < gB$, weak SO coupling
regime); (b) $\alpha = \beta$ = 0.1 ($\gamma^2 = gB$, intermediate
SO coupling regime); (c) $\alpha = \beta$ = 0.2 ($\gamma^2 > gB$,
strong SO coupling regime).

For the case of weak SO coupling regime shown in
\fig{fig4.3.4.1}(a), the attractive scattering potential may induce
a Fano structure in conductance. This is because a true-bound-state
can be induced by the attractive scattering potential at energy $E =
E_1^+ - V_0^2/4$, in which the binding energy $E_b$ = $V_0^2/4$ =
$0.0025$ (dashed), $0.01$ (dash-dotted), and $0.0225$ (solid).  The
Fano structure is at $E\approx 0.99 E^\ast$ for potential $V_0 =
-0.3$. It is interesting to notice that the bounded upper
spin-branch electron bounded energy interfere with the extended
lower spin-branch electron and form the RD-Zeeman induced Fano
structure. For the case of intermediate SO coupling regime shown in
\fig{fig4.3.4.1}(b), we can see clear quasi-bound-state feature at
the subband bottom of the upper spin branch.  For the case of strong
SO coupling regime shown in \fig{fig4.3.4.1}(c), the Fano structure
is red-shifted to $E\approx 0.91 E^\ast$ for potential $V_0 = -0.3$.
In the SOI-Zeeman induced subband gap region, we see a more
significant quasi-bound-state formed at around $E\approx 1.02
E^\ast$.

\begin{figure}[h]
\includegraphics[width=0.4\textwidth,angle=0] {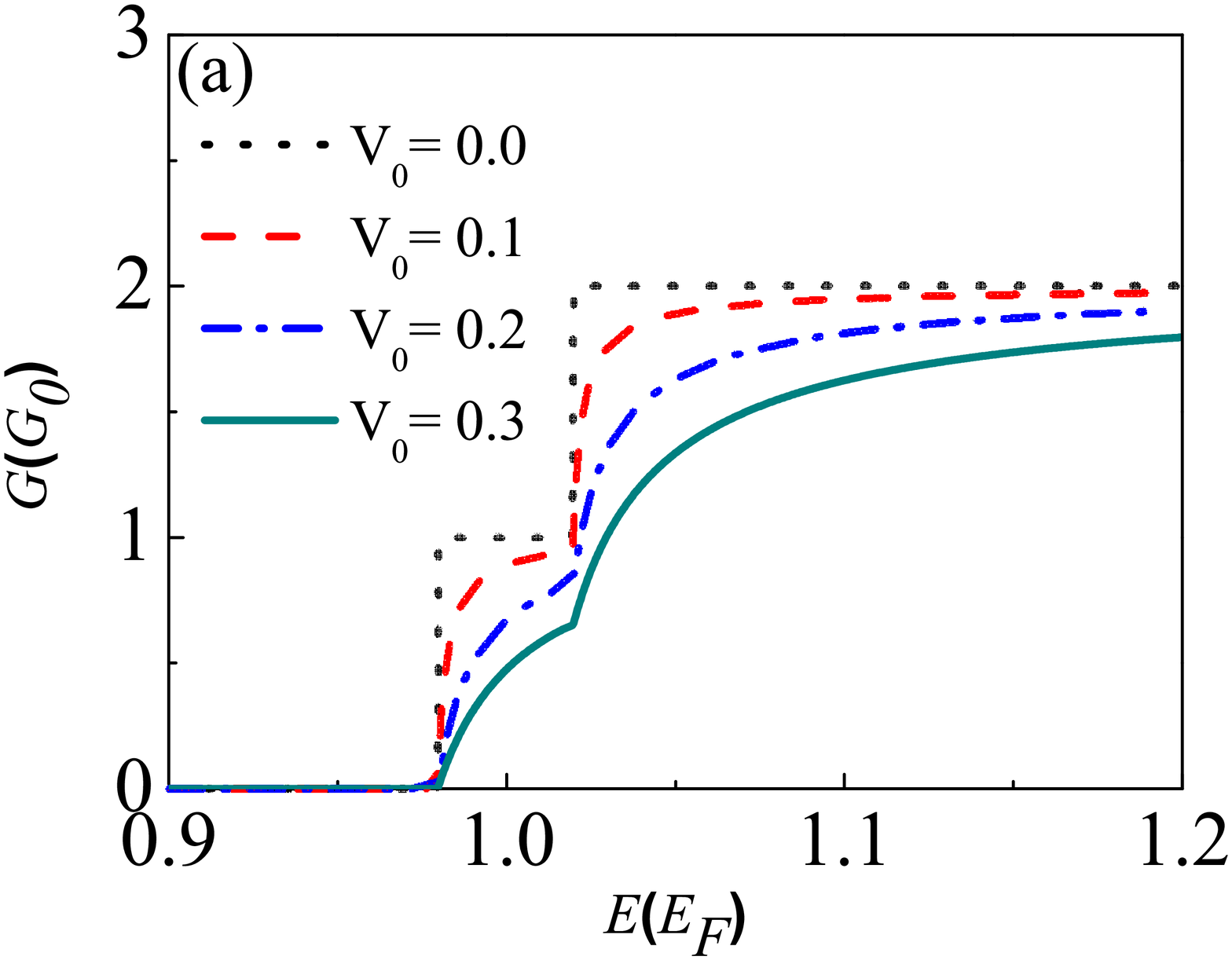}
\includegraphics[width=0.4\textwidth,angle=0] {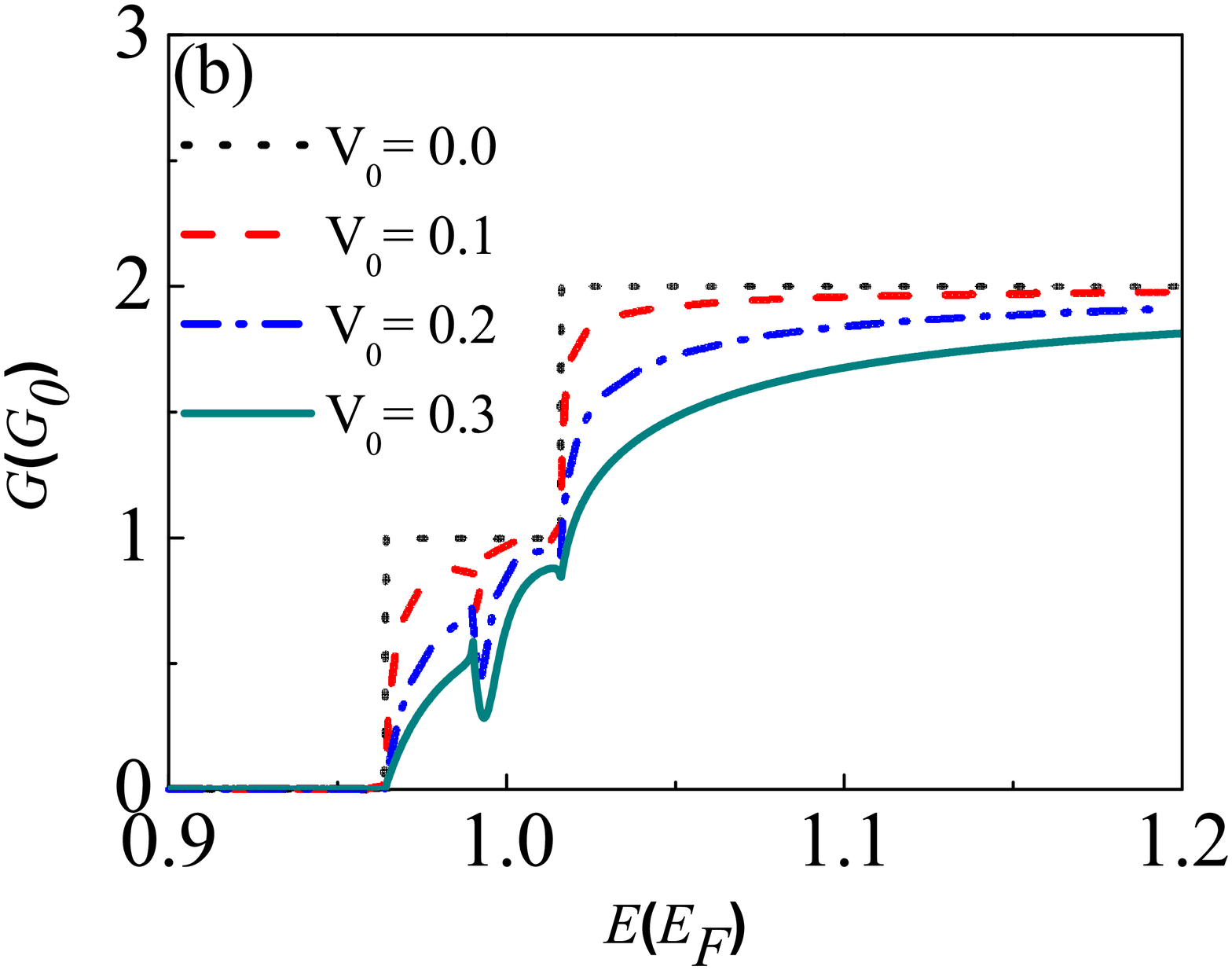}
\includegraphics[width=0.4\textwidth,angle=0] {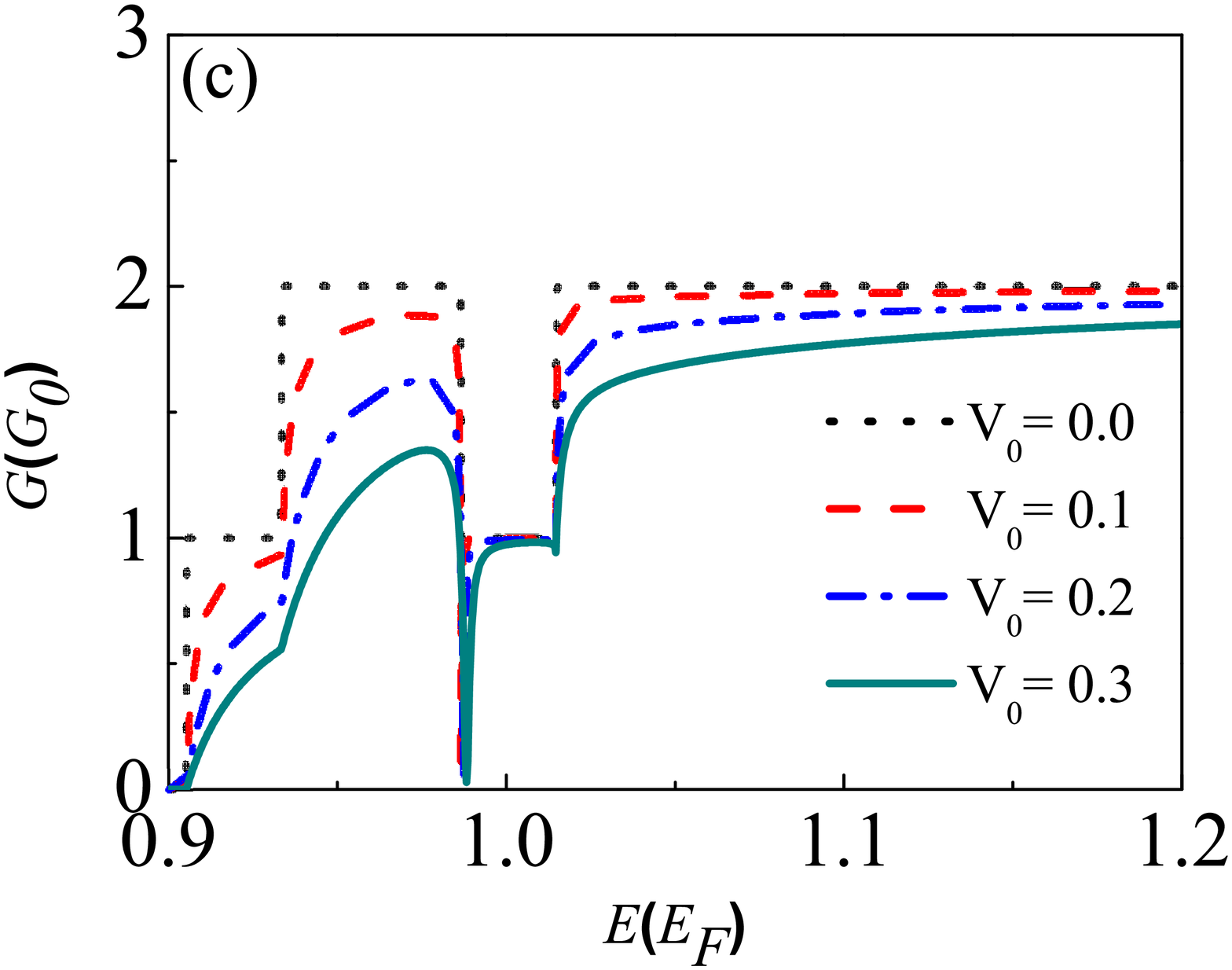}
\caption{(Color online) Conductance as a function of electron energy
with different strength of repulsive scattering potential: $V_0 =
0.0$ (dotted black), $V_0 = 0.1$ (dashed red), $V_0 = 0.2$
(dash-dotted blue), $V_0 = 0.3$ (solid green). We fix the in-plane
magnetic field ($gB$ = 0.02) and tune the Rashba and Dresselhaus
SO-coupling constants: (a) $\alpha = \beta$ = 0.02 ($\gamma^2 < gB$,
weak SO coupling regime); (b) $\alpha = \beta$ = 0.1 ($\gamma^2 =
gB$, intermediate SO coupling regime); (c) $\alpha = \beta$ = 0.2
($\gamma^2 > gB$, strong SO coupling regime).} \label{fig4.3.4.4}
\end{figure}

In \fig{fig4.3.4.4}, we investigate how a repulsive scattering
potential influences the transport properties by tuning Rashba and
the Dresselhaus effects and fixing the in-plane magnetic field, the
corresponding energy spectra are shown in \fig{fig4.1.3.1}. The
conductance is plotted as a function of electron energy with
different strength of repulsive scattering potential: $V_0 = 0.0$
(dotted), $V_0 = 0.1$ (dashed), $V_0 = 0.2$ (dash-dotted), $V_0 =
0.3$ (solid). We fix the in-plane magnetic field so that the Zeeman
effect $gB$ = 0.02. In addition, we tune the Rashba and Dresselhaus
SO-coupling constants as (a) $\alpha = \beta$ = 0.02 ($\gamma^2 <
gB$, weak spin-orbit coupling regime); (b) $\alpha = \beta$ = 0.1
($\gamma^2 = gB$, intermediate spin-orbit coupling regime); (c)
$\alpha = \beta$ = 0.2 ($\gamma^2 > gB$, strong spin-orbit coupling
regime).

For the case of weak SO coupling regime shown in
\fig{fig4.3.4.4}(a), the repulsive scattering potential cannot form
bound states even for the case of strong potential amplitude $V_0 =
0.3$, in which the conductance behaves monotonically increasing and
the conductance is suppressed to $G \approx 0.7 G_0$.  For the case
of intermediate SO coupling regime shown in \fig{fig4.3.4.4}(b), it
is interesting to note that the conductance manifests a hole-like
quasi-bound-state feature on the top of shoulder subband top ($E =
0.99 E^\ast$), as is  shown in \fig{fig4.1.3.1}(b). For the case of
strong SO coupling regime shown in \fig{fig4.3.4.4}(c), the general
feature in conductance is the strong drop from $2G_0$ to $G_0$ in
the subband gap of the two spin branches.  Moreover, it is clearly
shown that the hole-like quasi-bound-state feature can be induced on
the subband top of the lower spin branch and form a very clear dip
structure in conductance.

\section{Concluding Remarks}

We consider a narrow constriction with the Rashba and Dresselhaus
spin-orbit interactions under an in-plane magnetic field applied in
the transport direction.  A top finger gate is used to generate an
attractive or a repulsive scattering potential.  This allows us to
investigate the coherent quantum transport properties involving
spin-flip scattering. The competition of the spin-orbit scattering
and the Zeeman effect plays an important role to the subband
structures and the transport properties.  The Zeeman effect allows
us to separate the R-SOI and RD-SOI into three regimes: the weak,
mediate, and strong SO coupling regimes.

In the weak SO coupling regime with Zeeman effect, the subband
structure remains the quadratic form. It is symmetric if only the
Rashba SOI dominates while asymmetric if both the Rashba and
Dresselhaus SOIs are significant.  For the case of attractive
potential with only the Rashba SOI, it allows electron occupying the
upper spin branch to form a true-bound-state feature with binding
energy $V_0^2/4$, and the conductance manifests a valley structure.
It is important to note that the presence of both the Rashba and the
Dresselhaus SOIs may enhance the interference between the localized
upper spin branch state and the extended lower spin branch state,
and hence the conductance manifests a Fano structure.  For the case
of repulsive potential the conductance behaves monotonically
increasing for both R-SOI and RD-SOI.

In the intermediate SO coupling regime with Zeeman effect, the
subband structure of the lower spin branch exhibits a quadratic
structure for R-SOI and a shoulder-like structure for RD-SOI.  For
the case of attractive potential with R-SOI, the conductance
manifests a quasi-bound-state feature below the upper branch.
Moreover, for the case of attractive potential with RD-SOI, we have
found a kink structure in conductance at the shoulder of the lower
spin branch. For the case of repulsive potential with R-SOI, the
conductance is strongly suppressed and monotonically increasing.
However, for the case of repulsive potential with RD-SOI, the
conductance can manifest a clear hole-like quasi-bound-state
feature.

In the strong SO coupling regime with Zeeman effect, the subband
structure of the lower spin branch exhibits a subband top structure
for both the R-SOI and RD-SOI.  In addition, the two subband bottoms
of the lower spin branch with same energy for R-SOI and with
different energy for RD-SOI. For the case of attractive potential
with R-SOI, the conductance manifests a quasi-bound-state feature
below the upper branch. However, the conductance structure for the
case of attractive potential with RD-SOI is more complicated.  We
have found a true-bound-state feature in conductance with Fano line
shape depending on the strength of scattering potential. This
behavior is due to the different energy of two subband bottoms in
the lower spin branch.  In addition, an electron-like
quasi-bound-state can be found at the subband bottom of the upper
spin branch. For the case of repulsive potential with R-SOI, we have
found clear hole-like quasi-bound-state feature at the subband top
of the lower spin branch. This hole-like quasi-bound-state feature
is more significant with longer life time for the case of repulsive
potential with RD-SOI.

In conclusion, we have investigated the interplay of
Rashba-Dresselhaus spin-orbit interaction and the in-plane magnetic
field induced Zeeman effect to influence the spin-resolved coherent
transport. By tuning the finger gate, we have demonstrated how the
attractive and repulsive scattering potentials affect the
conductance features. We have analyzed in detail the nontrivial
subband and quantum transport properties concerning the SOI-Zeeman
induced electron-like and hole-like quasi-bound-state features.

%-----------------------------------------------------------------------------
%
%
\begin{acknowledgments}
This work was supported by the National Science Council in Taiwan
under Grants No.\ NSC100-2112-M-239-001-MY3,  No.\
NSC-98-2112-M-009-011-MY2, and No.\ NSC-100-2112-M-009-013-MY2. We
are thankful to the technical support from Shu-Jui Yu.

\end{acknowledgments}
%
%---------------------------------------------
%

\end{document}